\documentclass[twocolumn]{aastex63}

\usepackage{xspace}

\usepackage{hyperref}
\def\iso#1{$^{#1}$}
\def\msun{M$_\odot$}

\received{June 4, 2021}
\revised{December 8, 2021}
\accepted{ecember 21, 2021}
\submitjournal{ApJ}

\shorttitle{iMAC}

\graphicspath{{./}{figures/}}

\begin{document}

\title{Comparison between core-collapse supernova nucleosynthesis and meteoric stardust grains: investigating magnesium, aluminium, and chromium}


\correspondingauthor{Maria Lugaro}
\email{maria.lugaro@csfk.org}

\author[0000-0003-1976-9947]{Jacqueline den Hartogh}
\affiliation{Konkoly Observatory, Research Centre for Astronomy and Earth Sciences, E\"otv\"os Lor\'and Research Network (ELKH), Konkoly Thege Mikl\'{o}s \'{u}t 15-17, H-1121 Budapest, Hungary}
\affiliation{NuGrid Collaboration, \url{http://nugridstars.org}}

\author[0000-0000-0000-0000]{Maria K. Pet{\"o}}
\affiliation{Konkoly Observatory, Research Centre for Astronomy and Earth Sciences, E\"otv\"os Lor\'and Research Network (ELKH), Konkoly Thege Mikl\'{o}s \'{u}t 15-17, H-1121 Budapest, Hungary}

\author[0000-0002-1609-6938]{Thomas Lawson}
\affiliation{Konkoly Observatory, Research Centre for Astronomy and Earth Sciences, E\"otv\"os Lor\'and Research Network (ELKH), Konkoly Thege Mikl\'{o}s \'{u}t 15-17, H-1121 Budapest, Hungary}
\affiliation{NuGrid Collaboration, \url{http://nugridstars.org}}
\affiliation{E.~A.~Milne Centre for Astrophysics, Department of Physics and Mathematics, University of Hull, HU6 7RX, United Kingdom}
\affiliation{Joint Institute for Nuclear Astrophysics - Center for the Evolution of the Elements}

\author[0000-0001-8235-5910]{Andre Sieverding}
\affiliation{School of Physics and Astronomy,
      University of Minnesota, Minneapolis, MN 55455, USA} 
\affiliation{Physics Division, Oak Ridge National Laboratory, P.O. Box 2008, Oak Ridge, TN 37831-6354, USA}      
      
\author[0000-0000-0000-0000]{Hannah Brinkman}
\affiliation{Konkoly Observatory, Research Centre for Astronomy and Earth Sciences, E\"otv\"os Lor\'and Research Network (ELKH), Konkoly Thege Mikl\'{o}s \'{u}t 15-17, H-1121 Budapest, Hungary}
\affiliation{Graduate School of Physics, University of Szeged, Dom t\'er 9, Szeged, 6720 Hungary}

\author[0000-0000-0000-0000]{Marco Pignatari}
\affiliation{Konkoly Observatory, Research Centre for Astronomy and Earth Sciences, E\"otv\"os Lor\'and Research Network (ELKH), Konkoly Thege Mikl\'{o}s \'{u}t 15-17, H-1121 Budapest, Hungary}
\affiliation{E.~A.~Milne Centre for Astrophysics, Department of Physics and Mathematics, University of Hull, HU6 7RX, United Kingdom}
\affiliation{NuGrid Collaboration, \url{http://nugridstars.org}}
\affiliation{Joint Institute for Nuclear Astrophysics - Center for the Evolution of the Elements}

\author[0000-0002-6972-3958]{Maria Lugaro}
\affiliation{Konkoly Observatory, Research Centre for Astronomy and Earth Sciences, E\"otv\"os Lor\'and Research Network (ELKH), Konkoly Thege Mikl\'{o}s \'{u}t 15-17, H-1121 Budapest, Hungary}
\affiliation{ELTE E\"{o}tv\"{o}s Lor\'and University, Institute of Physics, Budapest 1117, P\'azm\'any P\'eter s\'et\'any 1/A, Hungary}
\affiliation{School of Physics and Astronomy, Monash University, VIC 3800, Australia}

\begin{abstract}
Isotope variations of nucleosynthetic origin among Solar System's solid samples are well documented, yet the origin of these variations is still uncertain. The observed variability of \iso{54}Cr among materials formed in different regions of the proto-planetary disk has been attributed to variable amounts of presolar chromium-rich oxide (chromite) grains, which exist within the meteoritic stardust inventory and most likely originated from some type of supernova explosions. To investigate if core-collapse supernovae (CCSNe) could be the site of origin of these grains, we analyse yields of CCSN models of stars with initial mass 15, 20 and 25 M$_{\odot}$, and solar metallicity. We present an extensive abundance data set of the Cr, Mg, and Al isotopes as a function of enclosed mass. We find cases in which the explosive C-ashes produce a composition in good agreement with the observed \iso{54}Cr/\iso{52}Cr and \iso{53}Cr/\iso{52}Cr ratios as well as the \iso{50}Cr/\iso{52}Cr ratios. Taking into account that the signal at atomic mass 50 could also originate from \iso{50}Ti, the ashes of explosive He-burning also match the observed ratios. Addition of material from the He ashes (enriched in Al and Cr relative to Mg to simulate the make-up of chromite grains) to the Solar System composition may reproduce the observed correlation between Mg and Cr anomalies, while material from the C-ashes does not present
significant Mg anomalies together with Cr isotopic variations.
In all cases, non-radiogenic, stable Mg isotope variations dominate over the variations expected from \iso{26}Al.

\end{abstract}



\section{Introduction} 
\label{sec:intro}

Isotopic differences of nucleosynthetic origin are observed among meteorite groups and primitive meteorite components that formed in the Solar System. For example, spinel-hibonite spherules and `normal' calcium and aluminum rich inclusions (CAIs) do not show nucleosynthetic variability, while ultrarefractory platy hybonite crystals and CAIs with fractionation and unidentified nuclear effects (also known as FUN CAIs) do. This is interpreted as a record of progressive homogenisation of dust and gas in the inner regions of the proto-planetary disk via turbulent mixing and thermal heating during the T-Tauri phase of the Sun \citep{2014EPSLMishra,2018ApJPignatale,2019ApJPignatale,2019ApJJacquet}. Nucleosynthetic isotope variations are also observed among bulk compositions of meteorites and planetary objects which implies that large scale isotopic heterogeneities, inherited from the proto-solar nebula and/or formed during the evolution of the proto-planetary disk, have been preserved. 

These variations, however, are hard to connect to nucleosynthetic signatures from specific stellar sources. A number of scenarios have been developed to explain such connection. These range from isotopic differences inherited from an inhomogeneous molecular cloud \citep{2019GeCoABurkhardt,2002ApJDauphas,2019EPSLNanne}, late processes acting on a once homogenized material in the inner regions of the proto-planetary disk \citep{2012EPSLBurkhardt,2017EPSLPoole,2009SciTrinquier,2008ApJDauphas,2008EPSLRegelous}, and/or new material added to the proto-planetary disk after the formation of the Sun \citep{2016PNASVanKooten, 2018NaturSchiller}.

Solids from the proto-planetary disk not only display variation in bulk isotopic compositions, but often also display a discontinuity (gap). For the isotopes of many elements (e.g. Cr, Ti, Mo, Ru), meteorite types are well separated into two groups. Because of this compositional gap, nucleosynthetic isotope variations are often called the ``isotopic dichotomy'' of the proto-planetary disk \citep{2011EPSLWarren}. Materials assumed to have formed in the outer Solar System are associated with enrichment in neutron-rich isotopes of intermediate-mass and iron group elements, such as $^{48}$Ca, $^{50}$Ti , $^{54}$Cr \citep[see e.g.][]{2007ApJTrinquier,2009SciTrinquier, 2018NaturSchiller}, neutron-capture affected isotopes such as those of Mo and Ru \citep[see, e.g.][]{2016EPSLBudde,2017PNASKruijer,2019EPSLNanne}, and other isotopes of explosive nucleosynthesis origin such as \iso{58}Ni \citep{2019EPSLNanne} and \iso{92}Nb \citep{hibiya19}, as compared to materials assumed to have formed in the inner Solar System\footnote{Material from the outer and inner Solar System are represented respectively by (i) carbonaceous chondrites and ``carbonaceous type'' iron meteorites, collectively referred to as CC; and (ii) ordinary chondrites, lunar and martian samples, ``non-carbonaceous'' iron meteorites and various achondrites, collectively referred to as NC.}, see, e.g., \citet{2016EPSLBudde,2017PNASKruijer,2019EPSLNanne,hibiya19} and 
the review by \citep{kleine20}.
 
The nucleosynthetic source of these enrichments has been attributed to supernovae but the exact origin is still unclear \citep[e.g.,][]{1985ApJHartmann, 2010ApJDauphas}. The formation of Jupiter's core \citep{2014Helled, 2017PNASKruijer}, or a pressure maximum in the disk leading to such formation \citep{brasser20} have been invoked as the barrier that kept these two reservoirs well separated in the early Solar System. 

The chromium isotopes are exceptionally useful to deconvolve the origin of planetary scale nucleosythetic isotope variation in iron group elements because Cr has four stable isotopes (at atomic mass 50, 52, 53, and 54), which allows us to obtain two ratios after mass-fractionation effects are removed with internal normalisation. Furthermore, it appears that the main feature of the Cr anomaly, i.e., enrichment and depletion of the most neutron rich isotope ($^{54}$Cr), is driven by a single, well-identified mineral carrier. \citet{2010ApJDauphas} and \citet{2010GCAQin} identified this carrier phase as Cr-oxide (with variable structure, but mostly chromium rich Mg-spinel, here referred to as chromite) and found that variable abundance of such presolar grains can explain all the variations observed among bulk meteorites. \citet{2018ApJNittler} provided high precision Cr data on these presolar chromite, confirming the previously assumed high \iso{54}Cr/\iso{52}Cr ratios (up to 80 times the solar ratio). \citet{2018ApJNittler} compared their data to a limited number of supernova models and concluded that the observations are better explained by models of electron capture supernovae \citep{2013ApJWanajo} and rare, high density type Ia SNe \citep{1997ApJWoosley} than by models of core collapse supernovae (CCSNe) by \citet{2007PhRWoosleyHeger}\footnote{Also in Asymptotic Giant Branch (AGB) stars neutron-capture processes can enrich the \iso{54}Cr relatively to the other Cr isotopes. However, the largest anomaly predicted in models of O-rich massive AGB stars does not exceed values in the order of 40\%, based on 6 \msun\ model of solar metallicity from \citep{karakas16}. Therefore, we can exclude that neutron captures in AGB stars are sources of presolar chromite with these anomalies.}. 

Interestingly, \iso{54}Cr variations among bulk meteorites and planetary objects may also correlate with mass independent $^{26}$Mg isotope variations (\citealt{2011ApJLarsen} and \citealt{2016PNASVanKooten}). The observed variation in $^{26}$Mg/$^{24}$Mg stable isotope ratio can be due to a heterogeneous distribution of the short lived radionuclide, $^{26}$Al (which decays to $^{26}$Mg with a half life of 0.717 Myr) along the proto-planetary disk, or to variations in stable $^{26}$Mg and/or $^{24}$Mg abundances, or both. \citet{2008Jacobsen} and \citet{Kita2013} argue for a homogeneous Mg isotope distribution in the Solar System. \citet{2011ApJLarsen} proposed that the apparent positive correlation between \iso{54}Cr and $^{26}$Mg anomalies among planetary objects is the result of progressive thermal processing of in-falling $^{26}$Al-rich molecular cloud material towards the inner regions of the disk. This in return results in preferential loss of thermally unstable and isotopically anomalous dust. Alternatively, \citet{2016PNASVanKooten} suggested that the apparent positive correlation between \iso{54}Cr and $^{26}$Mg may represent ``unmixing'' of distinct dust populations with different thermal properties. Old, thermally processed, presolar, homogeneous dust could mix with fresh, thermally unprocessed, supernova-derived dust, which formed shortly before the Solar System. This newly condensed dust is then preferentially lost from the inner regions of the proto-planetary disk.
Because of the high significance of this apparent correlation and its possible interpretations, we also make a first attempt to address it here from the point of view  of stellar modelling by exploring the Al and Mg isotopic composition of the specific CCSN regions that match the nucleosynthesis anomalies in presolar chromite grains. With simple mixing relations, we investigate if these CCSN abundances can generate any significant variation in $^{26}$Al or stable Mg isotopes among planetary objects. This is a simplified first attempt to linking stardust data to meteorites and planetary objects because it assumes that such Al and Mg abundances are carried in the chromite grains and/or similar carriers enriched in Al. While this is a possible scenario, there is no evidence for it yet as there are no Mg or O isotope studies on chrmomite presolar grains.

Here, we compare the predictions from three sets of CCSN models, from stars of initial mass 15, 20, and  25 M$_{\odot}$ and solar metallicity, to the chromite data to evaluate the role of CCSNe as potential sources of chromite grains in the large scale heterogeneity of the proto-planetary disk. We will also compare the abundances of the stable isotopes of Cr, Al, and Mg and \iso{26}Al in three sets of CCSN models and evaluate the isotopic abundances and ratios as a function of the enclosed stellar mass. Our aims are: first, to identify \iso{54}Cr production sites within CCSNe that may match the chromite grains; second, to evaluate the \iso{26}Al production and Mg isotope compositions associated with such \iso{54}Cr production sites; and third, to investigate if the Al and Mg abundances of the CCSN region of potential origin of the chromite grains could produce variation in \iso{26}Al or $^{26}$Mg isotopes among planetary objects (under the simple assumption described above).
Furthermore, we compare the \iso{26}Al production in the CCSN models to the \iso{26}Al signatures in the presolar grains from CCSN that we found in the literature, in order to put our analysis of potential Al and Mg abundances in chromite grains into the wider context of CCSN stardust grains in general. 

The structure of the paper is as follows: in Section \ref{sec:methods} we briefly describe the specifics of the CCSN data sets and outline their differences. The comparison of the total yields for the nine Al, Mg, and Cr isotopes is presented in Section \ref{sec:resultYields}, and in Section \ref{sec:resultsGrains} we present the comparison between the CCSN models and the observed Cr isotopic compositions of stardust grains, as well as the comparison of the CCSN models to \iso{26}Al/\iso{27}Al in other presolar CCSN grains. In our discussion in Section \ref{sec:discussion} we present the effects of uncertainties associated to neutron-capture reaction rates, an analysis on the Al and Mg isotopic composition of the \iso{54}Cr production sites, and a comparison between modeled ejecta compositions and the meteoritic data. Our conclusions are presented in Section \ref{sec:conclusions}. 

\section{Methods}
\label{sec:methods}

Core-collapse supernovae (CCSNe) are explosions associated with the death of massive stars that process their initial composition through a sequence of hydrostatic nuclear burning stages until an Fe core is formed \citep[see][for an extensive review]{Langer2012}. The self-consistent modeling of the explosion mechanism is still challenging and requires three-dimensional, high-resolution simulations, which are currently too expensive to allow us to perform large-scale surveys for nucleosynthesis studies \citep[for recent reviews see][]{2013RMPBurrows,2016PASAMueller,2016ARNPSJanka}. Instead, parameterized, spherically symmetric simulations have been employed widely to estimate CCSN nucleosynthesis yields. In such models, the innermost part of the CCSN progenitor is usually not simulated in detail but replaced with an engine that artificially drives the explosion, such as a piston \citep{1995ApJSWoosley} or the injection of thermal energy \citep{2003ApJLimongi}, which can be tuned with a few model-specific parameters to yield a desired explosion energy, measured as the kinetic energy at infinity, and remnant mass, which is referred to as the mass cut. In addition to the neutron star that is left behind by the explosion, the mass cut also includes the possibility of fallback, i.e., material that is initially ejected, but remains gravitationally bound to the remnant and thus eventually falls back onto it, possibly leading to the formation of a black hole even after a successful explosion \citep{2008ApJZhang,2009ApJFryer}. 
Recently, models have been developed that treat the evolution of the stellar core in more detail, instead of with an engine, and still achieve explosions in spherically symmetric simulations by different parameterizations \citep{2015ApJPerego,2016ApJSukhbold,2020ApJCouch}. Such models are promising to improve on the simple models mentioned above, but remain to be validated by comparison to multi-dimensional simulations and observations.

We collected three CCSN yield sets for $^{24,25,26}$Mg, $^{26,27}$Al and $^{50,52,53,54}$Cr \citep[Lawson et al. (submitted),][]{2018ApJSieverding,Ritter2018}, for which we have access to abundance profiles as a function of the stellar mass coordinate. These yield sets are based on 1D calculations using different stellar evolution, explosion, and post-processing codes. We include only models with an initial mass of 15, 20, and 25 M$_{\odot}$ at solar metallicity, since higher mass stars are expected to result in the formation of a black hole without any significant ejection of material processed by explosive nuclear burning \citep{2003ApJHeger}. We exclude the effects on yields by rotation, magnetic fields, and binary evolution as these are not known or too uncertain \citep{2019ARAAAerts,2019AAJacqueline,2020AABelczynski}. The yield sets are listed in Table~\ref{tab:yielddetails}, together with details on codes used for the calculations. For each data set we list in the following subsections the codes used for the calculations and the details of the initial set-ups that are important for our comparison. 

\subsection{Data set of Lawson et al. (submitted, LAW)}

The LAW models are a part of the large data set presented in \citet{2018ApJFryer}, who performed a parameter study over a broad range for SN explosions. \cite{Andrews2020} used these models to study the production of radioactive isotopes relevant for the next generation of facilities for $\gamma-$ray astronomy, and provided the complete yields for the full stellar set. \cite{Jones2019} used the same set to study the production of $^{60}$Fe. Here we use the updated yield set based on the same models, but updated by including a recent bug fix (Lawson et al., submitted).

The progenitor stellar evolution models were calculated with a recent version of the Kepler hydrodynamic code \citep{1978ApJWeaver,Heger2010}, using initial abundances based on \citet[][GN93]{1993Grevesse}. The progenitors were post-processed to obtain the detailed nucleosynthetic results using MPPNP (Multi-zone Post-Processing Network -- Parallel, see \citealt{Pignatari2016a} and \citealt{Ritter2018}). The explosions of the progenitors are calculated using a 1D code mimicking a 3D convective engine, as described in \cite{Herant1994} and \cite{Fryer1999}. The explosion nucleosynthesis is calculated using TPPNP \citep[Tracer particle Post-Processing Network -- Parallel, see][]{2019MNRASJones}. The difference between MPPNP and TPPNP is that the first also performs mixing of mass shells following the mixing as calculated in the progenitor or explosion model, while the latter does not apply any mixing and may efficiently streamline the post-processing of trajectories. The same nuclear reaction package is used from the two post-processing frameworks. 

\subsection{Data set of \citet[SIE]{2018ApJSieverding}}
\label{sec:sie}

The progenitor models of SIE were calculated with a slightly older version of the Kepler code than the LAW models. Differences include the neutrino loss rates as discussed by \citet{2018ApJSukhbold} and updated photon opacities. Due to these differences, the SIE models show a less massive C/O core and more compact structure than the LAW models. The initial abundances for the progenitors of SIE are based on \citet[][L03]{2003ApJLodders}. The explosion was simulated with a piston, as described in \citet{1995ApJSWoosley}. The piston is put at the mass cut determined by the position where the entropy per baryon drops below 4 $k_B$. The parameters of the piston were adjusted to produce an explosion energy of  $1.2\times 10^{51}\,\mathrm{erg}$. All matter outside the mass cut is assumed to be ejected, i.e., no additional fallback is considered. 

The explosive nucleosynthesis was post-processed by \citet{2018ApJSieverding}, who performed a parameter study around the effects of neutrino energies. We include here the models with the highest neutrino energy.

\subsection{Data set of \citet[RIT]{2018MNRASRitter}}

The progenitor models of \citet{2018MNRASRitter} were calculated with the MESA stellar evolution code \citep{2011ApJSPaxton} with initial abundances based on GN93. The explosion models were calculated via the semi-analytical approach using the delayed formalism as described in \citet{2016ApJSPignatari}, and using the mass cuts from \citet{2012ApJFryer}. The detailed nucleosynthesis was calculated for the progenitor and the explosion with the post-processing code MPPNP. 
In the evolution of the 15 M$_{\odot}$ star the convective O and C shells merge. This feature can occur during the later phases of stellar evolution, when the different burning shells are formed close enough to each other to possibly merge. The shell merger in the 15 M$_{\odot}$ progenitor model takes place at the end of the core Si burning phase, see Appendix \ref{sec:app} for more details. Shell mergers are often found in 1D and 3D stellar evolution models \citep[see][for a recent review]{2020LRCAmuller}, and shell-merger events are often initiated shortly before the collapse. \citet{2018MNRASCollins} found that 40\% of their stellar evolution models with an initial mass between 16 and 26 M$_{\odot}$ start the core collapse during an ongoing shell-merger. 

\subsection{Decayed abundances}
\label{sec:dec_abu}
We present the isotope abundances and isotopic ratios as a function of stellar mass coordinates for both the progenitor and explosion models. Unless indicated otherwise, in the following figures we show the abundances obtained after decaying all the radioactive isotopes created during the explosion into their respective stable isotope, except for the case of \iso{26}Al, as we want to study its production. For the isotopes of interest here, the most relevant decay chain is  \iso{53}Mn($\beta^+$)\iso{53}Cr with a half life of 3.74 Myr. Its effect on the comparison to the stardust grains will be considered in Section~\ref{sec:chromite}.

\begin{table*}
    \centering
    \caption{Overview of the details of the different yield sets included in this paper. The first three studies are discussed in detail in this work, the last four are only included here and in Section \ref{sec:comp_yields}. }    
    \begin{tabular}{r|c|c|c|c}
       Set & Code for progenitors & Code for explosions & Initial abundances & Solar metallicity\\
       \hline
       Lawson et al.(submitted) (LAW) & Kepler & Convective engine & GN93 & Z=0.02  \\
       \citet{2018ApJSieverding} (SIE) & Kepler & Kepler & L03 & Z=0.013  \\
       \citet{2018MNRASRitter} (RIT) & MESA & Semi-analytical & GN93 & Z=0.02  \\
       \hline
       \citet{2018ApJSLimongi}\footnote{This study also investigates the effects of rotation, but we exclude those models in our comparison due to the large uncertainties present in the theory of rotation in stellar evolution \citep[see e.g.][]{2019ARAAAerts,2020AABelczynski,2019AAJacqueline}.} (LIM) & FRANEC & FRANEC & AG89\footnote{\citet{1989GeCoAAnders}\label{AG89}}& Z=0.02 \\
       
       \citet{Rauscher2002} (RAU) & Kepler\footnote{Progenitor models from \citet{1995ApJSWoosley} (all other Kepler progenitor models are more recent)} & Kepler & AG89\textsuperscript{\ref{AG89}}& Z=0.02 \\
       
       \citet{2019ApJCurtis} (CUR)  & Kepler & PUSH & L03 & Z=0.013  \\
       
       \citet{2016ApJSukhbold}\footnote{We include two models (14.9 and 25.2 M$_{\odot}$) as shown in the paper, other yields can be found in their online data.} (SUK) & Kepler & Kepler (W18 engine) & L03 & Z=0.013\\
    \end{tabular}
    \label{tab:yielddetails}
\end{table*}

\subsection{Nomenclature}
\label{sec:nomen}
In the following sections we define the regions within the stellar model from the envelope towards the core in the following way:
\begin{itemize}
    \item If \iso{4}He is the most abundant isotope, the region is called \textit{H-ashes} (the grey band located at the highest mass coordinate in the three panels of Figure~\ref{fig:struc_plots});
    \item If \iso{12}C and \iso{16}O are the most abundant isotopes, the region is called \textit{He-ashes} (the white band located at the highest mass coordinate in the three panels of Figure~\ref{fig:struc_plots});
    \item If \iso{16}O and \iso{20}Ne are the most abundant isotopes, the region is called \textit{C-ashes} (the grey band located to the left of the He-ashes in the three panels of Figure~\ref{fig:struc_plots});
    \item If \iso{16}O and \iso{28}Si are the most abundant isotopes, the region is called \textit{Ne-ashes} (the white band located to the left of the C-ashes in the three panels of Figure~\ref{fig:struc_plots}). The shell merger region in the 15 M$_{\odot}$ RIT model is also labelled as Ne-ashes;
    \item If \iso{28}Si is the most abundant isotope, the region is called \textit{O-ashes} (the grey band located to the left of the Ne-ashes in the three panels of Figure~\ref{fig:struc_plots});
    \item If \iso{56}Ni and \iso{56}Fe are the most abundant isotopes, the region is called \textit{Si-ashes} (the white band located to the left of the O-ashes in the three panels of Figure~\ref{fig:struc_plots}).
\end{itemize}
This nomenclature represents a simplified structure of the regions within massive stars before the explosion, and is often used within the massive star community. The nomenclature of \citet{1995Meyer} is commonly used in the presolar grain community and, when we compare our version to theirs, identifies mostly the same zones. The main difference is that they name the zone based on the most abundant isotopes, while our names refer back to the main fuel within the region. When putting the two schemes next to each other we get: our H-ashes are their He/N and He/C zones, our He-ashes are their O/C zone, our C-ashes are their O/Ne zone, our Ne-ashes are their O/Si zone, our O-ashes are their Si/S zone, and our Si-ashes are their Ni-zone.

\section{Yields}
\label{sec:resultYields}

We discuss in the first subsection the creation of $^{24,25,26}$Mg, $^{26,27}$Al and $^{50,52,53,54}$Cr in massive stars. Our analysis is focused on these nine isotopes, and their distribution in CCSN ejecta. The total isotopic yields of the data sets are compared in the second subsection. 

\subsection{Creation of the Al, Mg, and Cr isotopes in massive stars and CCSNe}
\label{sec:creation_overv}

We trace the internal structure of the models by plotting the abundance profiles of the mass fractions of $^4$He, $^{12}$C, $^{16}$O, $^{20}$Ne, $^{28}$Si, $^{56}$Ni, and $^{56}$Fe as a function of mass coordinate. In Figure~\ref{fig:struc_plots} we show the structure plots of the 15 M$_{\odot}$ models of LAW, SIE, and RIT. In Appendix \ref{sec:app} we provide figures of all three initial masses and the three data sets, showing the internal structure and also the final mass fractions of the Mg, Al, and Cr isotopes of the progenitor and the explosion model\footnote{All the data used to produce the figures in this paper can be found as Supplemental Data in the online Article Data}. 

\begin{figure*}
    \centering
    \includegraphics[width=\linewidth]{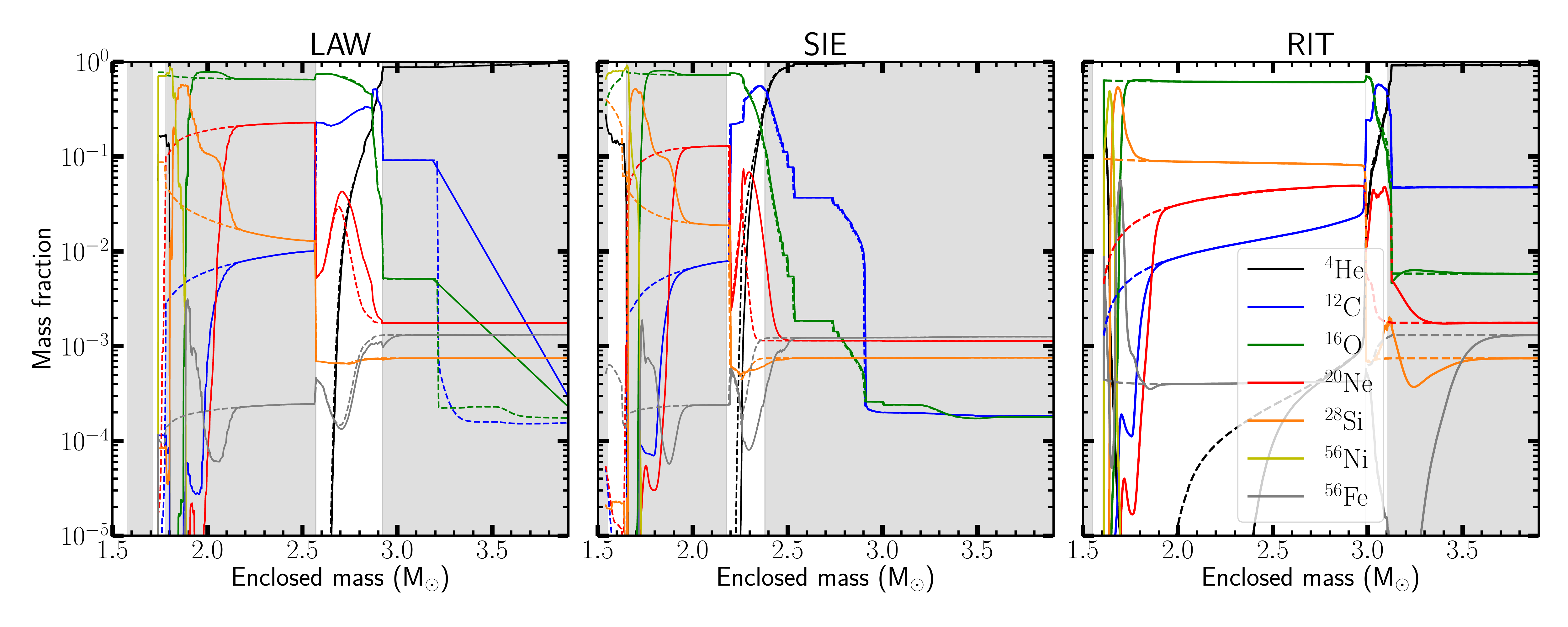}
    \caption{Structure plots showing the three (non-decayed) 15 M$_{\odot}$ models with grey and white bands to indicate the different regions (labelled as `ashes' in the text). The more compact structure of SIE compared to LAW is visible when comparing the mass coordinates of the bands. Furthermore, the RIT model shows a different internal structure due to the shell-merger. See Appendix \ref{sec:app} for the Al, Mg, and Cr isotope plots.}
    \label{fig:struc_plots}
\end{figure*}

\begin{table*}
    \centering
    \caption{The burning phases in the progenitor and explosion where the isotopes of interest are created and destroyed for the 15 M$_{\odot}$ LAW model. The burning phases are labelled using the definitions which were set in Section \ref{sec:nomen} (the word `ashes' is not repeated). The dominant nucleosynthesis process responsible for the production or destruction of each isotope are also indicated and explained in more detail in the text. (`cap' is short for `capture'). This table is specifically for 15 M$_{\odot}$, which might be where the differences with Table 3 of \citet{Woosley2002} come from. When comparing our table to \citet[][]{2019ApJCurtis}, the differences are due to our table excluding radiogenic contributions and including outer layers, which are the exact opposites of the features in the table of \citet{2019ApJCurtis}.}
    \begin{tabular}{l||l|l|l|l||l|l|l|l}
    & \multicolumn{4}{c||}{Progenitor} & \multicolumn{4}{c}{Explosion} \\
    \cline{2-9}
     & Produced in & Via & Destroyed in & Via & Produced in & Via & Destroyed in & Via  \\
    \hline
    \hline
    \iso{24}Mg & C, Ne, He & $\alpha$-cap & - & -     & He & $\alpha$-cap & O, Si & photo-dis \\
    \iso{25}Mg & C, Ne, He &  n-cap &Ne & photo-dis   & He & n,$\alpha$-cap & C & $\alpha$-cap\\
    \iso{26}Mg & C, Ne, He & n-cap &Ne & photo-dis    & He & n-cap & C & $\alpha$-cap\\
    \hline
    \iso{26}Al & C  & p-cap & Ne, He & several         & C & p-cap & O & photo-dis \\
    \iso{27}Al & C, Ne & p-cap &O & photo-dis         & - & - & O & photo-dis \\
    \hline
    \iso{50}Cr & C & p-cap & He & n-cap             & O, Si & p-cap & Si & equi \\
    \iso{52}Cr & - & - & He, C & n-cap              & O & equi & Si & equi \\
    \iso{53}Cr & C & n-cap & He & n-cap             & He & n-cap & C & n-cap \\
    \iso{54}Cr & He & n-cap & - & -                 & He & n-cap & C & n-cap \\
    \end{tabular}
    \label{tab:isotopes}
\end{table*}

Table \ref{tab:isotopes} shows the production and destruction sites of the nine isotopes of interest, plus their dominant reaction paths for the 15 \msun\ LAW model. Two reaction paths require explanation: `Photo-dis' stands for photo-disintegration, the process where an incoming photon removes a neutron, proton, or an $\alpha$-particle from the nucleus. `Equi' denotes the production in high temperature equilibrium conditions when most forward- and backward reaction rates are closely matches \citep{1973Woosley,1998ApJChieffi}. This usually applies to explosive Si and O burning.

In the following we highlight the most important differences between the models with respect to the production and destruction of the isotopes we are interested in. The 15 \msun\ SIE progenitor model shows production and destruction sites (Figure \ref{fig:M15_spagh}) that are comparable to the 15 \msun\ LAW model. The 15 \msun\ RIT progenitor model, however, experiences a shell-merger, which allows for C-burning while He-burning is still ongoing. Furthermore, the shell-merger allows for mixing of Cr-isotopes from the deeper layers outwards. In this shell-merger region, the 15 \msun\ RIT model shows a higher abundance for \iso{26}Mg than \iso{25}Mg (the opposite is visible in the LAW and SIE 15 \msun\ models), and the presence of \iso{50,52}Cr mixed up from deeper layers, which is not taking place in the LAW and SIE 15 \msun\ models. The 15 \msun\ explosive model of SIE and RIT show more explosive He burning than the 15 \msun\ LAW model. This allows for the production of the Mg-isotopes and extra destruction of \iso{50,52}Cr in the SIE and RIT 15 \msun\ models.

The main difference between the 15 and 20 \msun\ LAW progenitor models is that the mass cut is higher in the 20 \msun\ model (Figure \ref{fig:M20_spagh}), which leads to excluding the Ne-ashes from the ejecta. The 20 \msun\ SIE model is the only 20 \msun\ model including the Ne-ashes, where \iso{26}Al is produced. The 20 \msun\ RIT model shows a mass cut similar to the 20 LAW model and a production of \iso{26}Al in the C-ashes, like the 15 \msun\ RIT model. The main difference between the three 20 \msun\ models is that they show different amounts of explosive nucleosynthesis. The 20 \msun\ LAW explosive model shows no explosive nucleosynthesis involving the nine isotopes in Table~\ref{tab:isotopes}. The 20 \msun\ SIE model, however, shows explosive nucleosynthesis in the inner regions, producing \iso{26}Al and \iso{50,52}Cr. The 20 \msun\ RIT model shows explosive nucleosynthesis in the whole star, due to its high temperature compared to the other 20 \msun\ models (Figure \ref{fig:temps}). The explosive He-burning in this model is similar to the 15 \msun\ RIT model, while the explosive C-burning leads to the creation of \iso{26}Al and \iso{50,52}Cr and the destruction of \iso{27}Al and \iso{53,54}Cr.

The 25 \msun\ progenitor models (Figure \ref{fig:M25_spagh}) are similar to the 15 \msun\ progenitor models, except for the high mass cut in the upper C-ashes in the 25 \msun\ RIT model. The other two models show mass cuts below the Ne-ashes. In the 25 \msun\ explosive models we see explosion nucleosynthesis only in the inner regions of the LAW and SIE 25 \msun\ models, and not from explosive He-burning. This means most explosive contributions to the nine isotopes of interest are still present. In contrast, the 25 \msun\ RIT model only experiences explosive He-burning as its mass cut is too high to include the other regions. 

\subsection{Comparison of the total yields}
\label{sec:comp_yields}

\begin{figure*}
\centering
    \includegraphics[width=\linewidth]{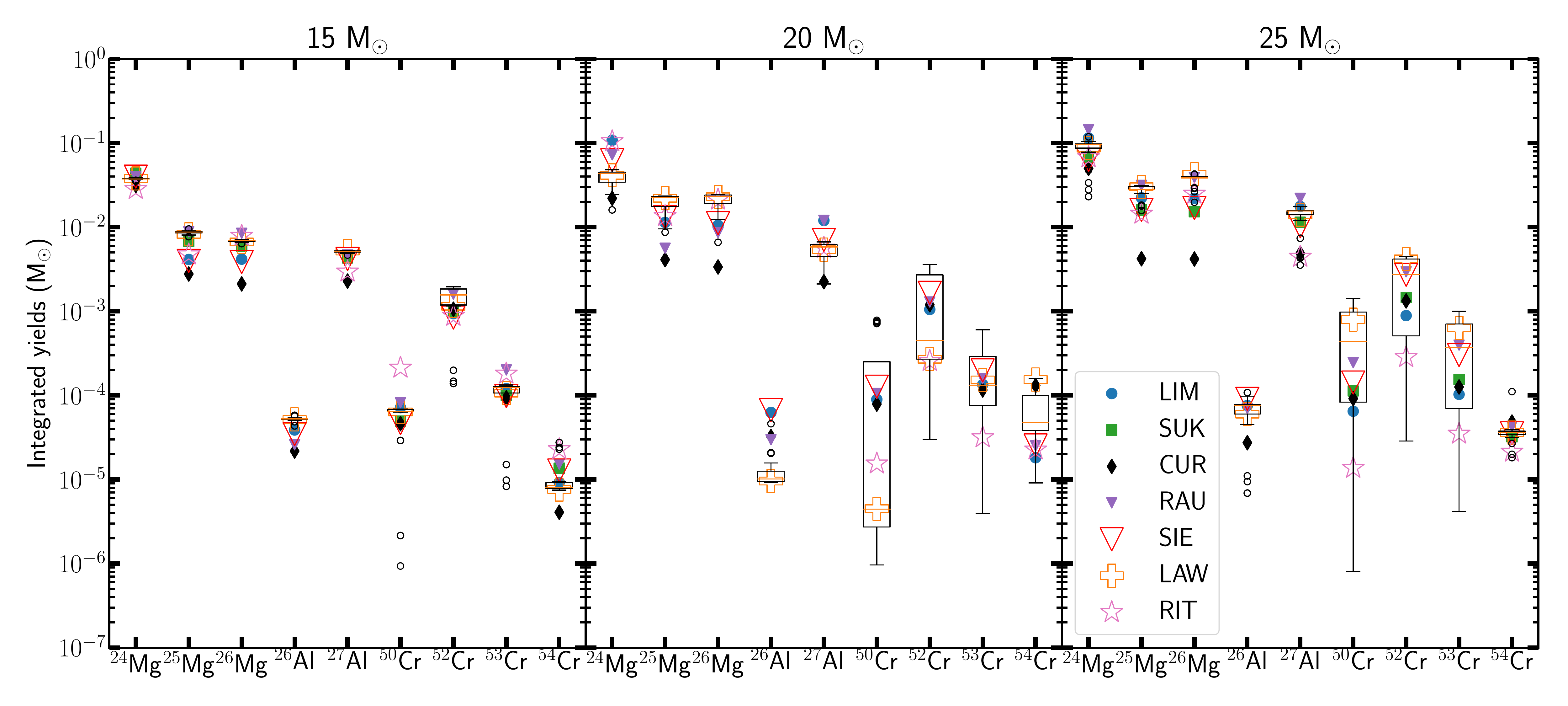}
    \caption{Comparison of the isotope yields of the three CCSN data sets considered in this paper, plus four others found in the literature. The three models of LAW that are considered in this paper and shown in the Appendix \ref{sec:app} are plotted as orange open crosses. The yields of all the models of LAW are included in the rectangle boxes, where the median value of the models is shown as orange line. The box plot covers the values between 25\% and 75\% of all the data points, while the error bars cover 1.5 times the range of the box plot. Any points outside the error bars are shown as open circles. The yield sets are labelled as indicated in Table~\ref{tab:yielddetails}.}
    \label{fig:isotope_boxplot}
\end{figure*}

Here we present a comparison of the total yields of seven CCSN data sets (not the net yields, which are calculated as the total yield minus the initial abundance). An overview of the main characteristics of the models by LAW, SIE, and RIT is given in Table~\ref{tab:yielddetails}, together with other four sets of CCSN models that are available in the literature \citep[][]{Rauscher2002,2016ApJSukhbold,2018ApJSLimongi,Curtis2018}. The seven sets have been calculated with different 1D stellar evolution and explosion codes. While this is not meant to be a comprehensive collection of CCSN yields available, it may be considered as indicative of the existing abundance variations obtained from different CCSN models. The comparison of the seven yields sets is presented in Figure~\ref{fig:isotope_boxplot}. We plot the explosive yields of the nine isotopes: $^{24,25,26}$Mg, $^{26,27}$Al and $^{50,52,53,54}$Cr, for all models grouped together according to their stellar masses. 

We note that the CUR yields are often the lowest yield for the Mg- and Al-isotopes. The reason for this is that this study only includes the inner stellar regions in their nucleosynthesis calculations. Parts of the C-ashes region are cut off, where the Mg and Al isotopes are abundant (see Table~\ref{tab:isotopes}), resulting in an apparent reduction of the total yield of the Mg and Al isotopes compared to other yields. Overall, we find that variations in the production of the nine isotopes in the seven yield sets are roughly one order of magnitude at most. The range of yields in the LAW models appears to cover most other yield sets, thus confirming that the parameter study of \citet[][]{2018ApJFryer} well represents the uncertainties within 1D CCSN explosions. We discuss in the remaining of this section the isotopes that show variations larger than one order of magnitude in the LAW, SIE, and RIT yields. 

The \iso{25,26}Mg and \iso{26,27}Al yields of LAW are higher than those of SIE in all panels of Figure~\ref{fig:isotope_boxplot}. This is because according to the LAW models the central stellar structures are less compact compared to SIE models (as mentioned in Section \ref{sec:sie}). Thus the Mg- and Al-rich C-ashes of LAW are located at higher mass coordinates than those of SIE.

Among the 15 \msun\ models (left panel), only the \iso{50,54}Cr yields show a spread of about one order of magnitude (excluding the few outliers of the LAW data set shown as small circles). The main reason for this is that the RIT model undergoes a shell-merger. In this region in the 15 \msun\ RIT model creates more Cr than the other two 15 \msun\ models, as the shell-merger transfers iron group elements from the deeper layers into the merged region (see Figure~\ref{fig:M15_spagh} and \citealt{2020ApJCote}). 

Among the 20 M$_{\odot}$ models (middle panel) again the Cr-isotope yields show the largest spread. The lower values of RIT are due to the higher mass cut values compared to the models of LAW and SIE (Figure~\ref{fig:M20_spagh}). This effect is not present in the Mg and Al isotopes, because these isotopes are produced in regions that are not affected by the mass cut. The large spreads in the models by LAW are caused by its large range of values for the mass cuts, see \citet{Fryer2018}. 

Also among the yields of the 25 \msun\ models (right panel) the Cr isotopes show the largest range of variations. The spread in the LAW data set is due to differences in the explosion energies. The Cr yields of RIT are again lower, due to its mass cut being higher than in models by LAW and SIE. 

In summary, the main differences between the three data sets of LAW, SIE, and RIT are the structural differences between the progenitors of the LAW and SIE data sets, the C-O shell merger in the 15 \msun\ RIT model, and the higher mass cuts in the 20 and 25 \msun\ RIT models.

\section{Results and comparison with presolar stardust grains}
\label{sec:resultsGrains}

Grains are formed locally within the CCSN ejecta, and thus we cannot use the total yields as presented in Section \ref{sec:comp_yields} for the comparison of CCSN yields to presolar chromite grains. Instead, we compare the high-precision grain data of \citet{2018ApJNittler} to the Cr isotopic ratios versus mass coordinate of the CCSN data sets of LAW, SIE, and RIT (Figures~\ref{fig:nittler_comp}-\ref{fig:nittler_comp50}). The ratios are plotted against the mass coordinates in Figure~\ref{fig:cr_ratios}. 
\citet{2018ApJNittler} also considered the possibility that the signal at atomic mass 50 represents \iso{50}Ti instead of \iso{50}Cr and report the \iso{50}Ti/\iso{48}Ti ratios inferred for 5 out of the 19 \iso{54}Cr-rich grains. Therefore, we also present and discuss here this possibility, while leaving the extended description of the production of the Ti isotopes in CCSNe to future work. 

Among the models of the LAW data set with different explosion energies, we use one for each initial mass in this section, which has an explosion energy closest to the value of $1.2\times 10^{51}\,\mathrm{erg}$ used by \citet{2018ApJSieverding}. The predicted isotopic ratios are calculated using decayed stellar abundances to consider the radiogenic contribution to the final abundances of stable isotopes (as explained in Section \ref{sec:dec_abu}), unless indicated otherwise. The boxes in Figures~\ref{fig:nittler_comp} and \ref{fig:nittler_comp50} are explained later in this section, when we precisely locate candidate regions that match the composition of the chromite grains.

We also explore the predicted Al and Mg isotope profiles of the CCSN models as a function of mass coordinate. In Figure~\ref{fig:groopman} we show the \iso{26}Al/\iso{27}Al ratio profiles of the CCSN models in comparison to the highest values determined for presolar grains of likely CCSN origin, such as SiC type X grains \citep[e.g.,][]{Groopman2015} and Group 4 presolar oxides \citep[e.g.,][]{2008ApJNittler}.

\subsection{Chromite grains}
\label{sec:chromite}
   
\begin{figure*}
    \centering
    \includegraphics[width=\linewidth]{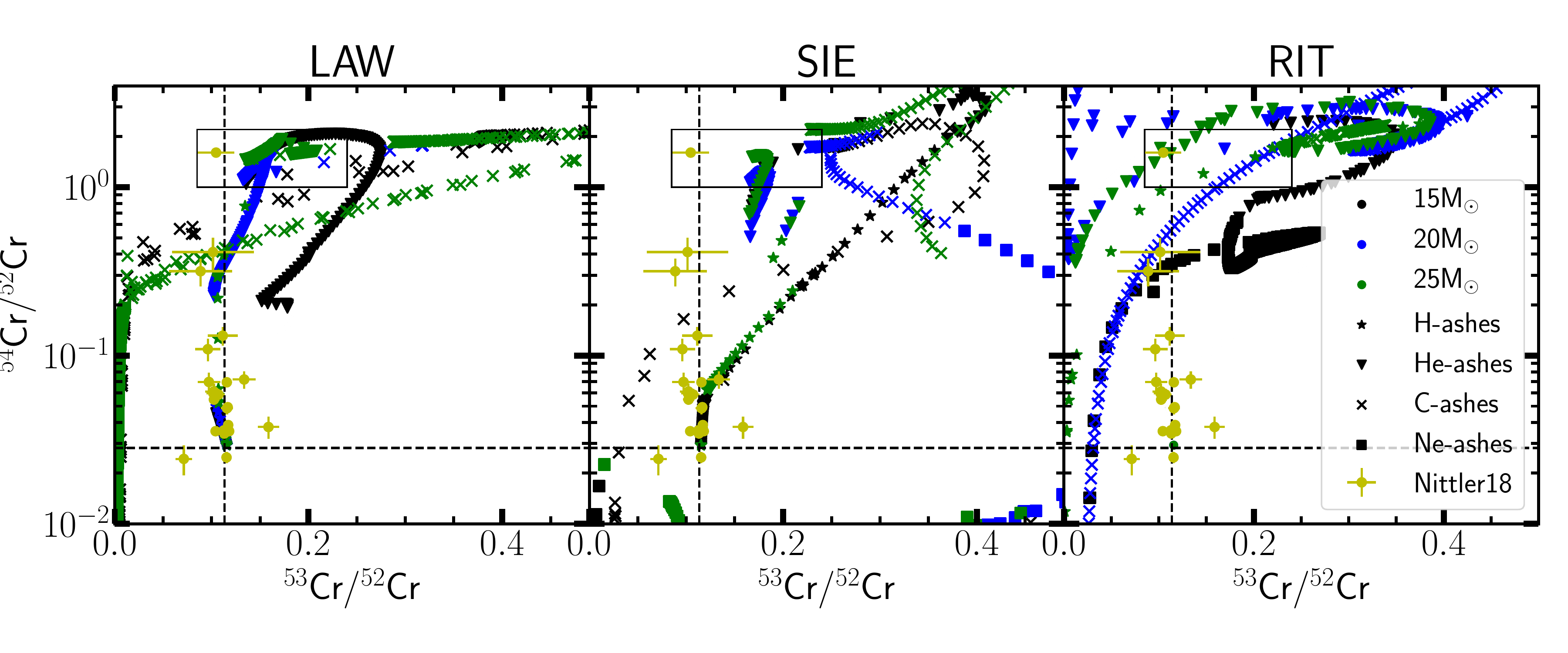}
    
    \vspace{-15pt}
    
    \includegraphics[width=\linewidth]{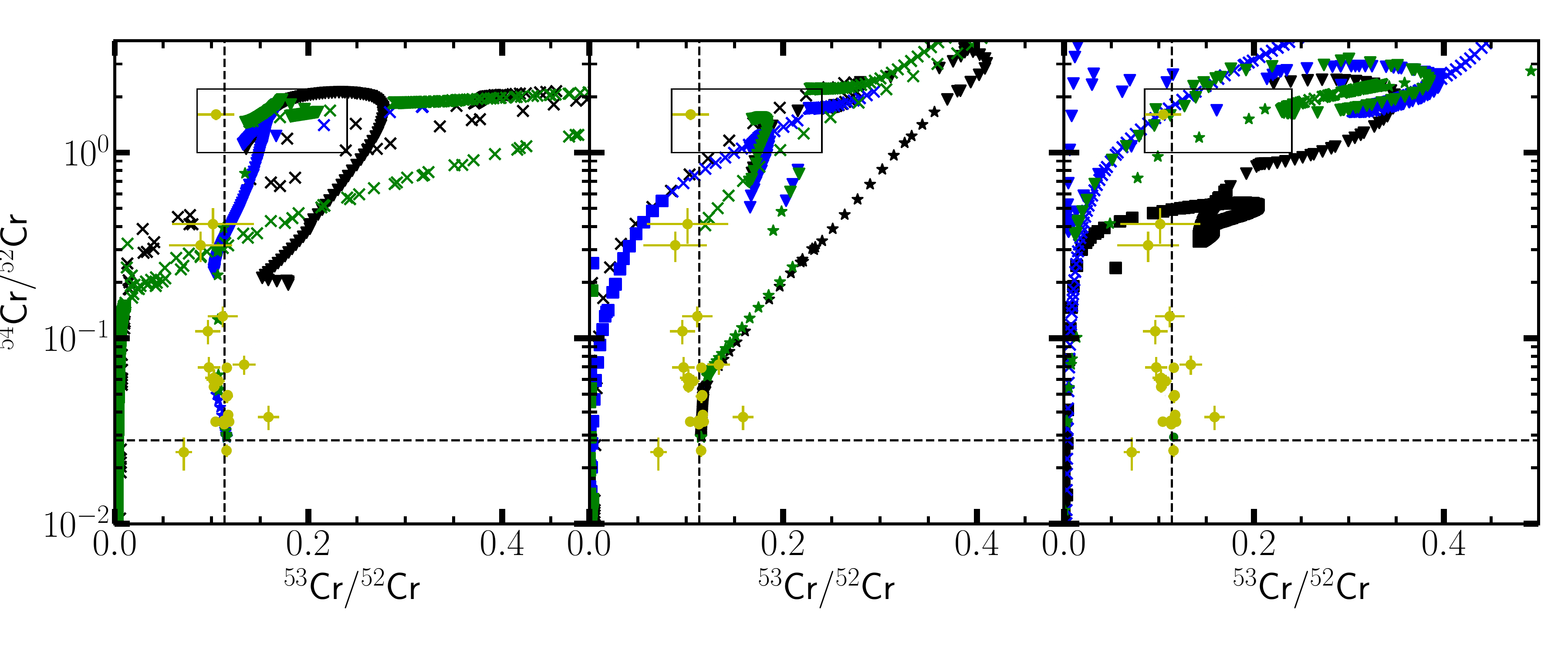}
    
    \vspace{-15pt}
    
    \includegraphics[width=\linewidth]{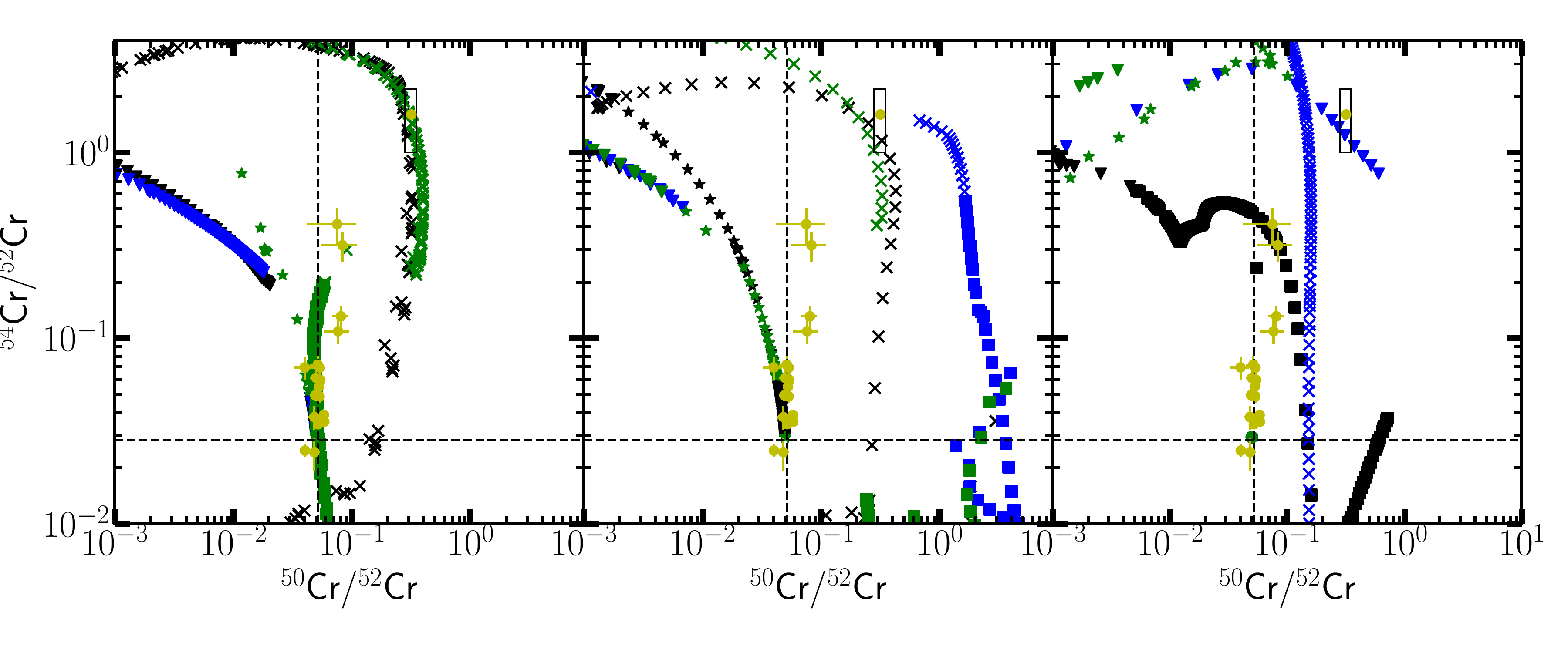}
    
    \vspace{-15pt}
    
    \caption{Comparison between the Cr isotopic ratios as measured by \citet{2018ApJNittler} (yellow data points with error bars), and those predicted by LAW, SIE, and RIT (left, middle, and right panels, respectively). Each point of the predictions corresponds to the composition of a mass shell, and different colors and symbols represent different initial masses and ashes, respectively (as indicated in the legend). Note that the symbols for the models are only plotted when the mass regions is O-rich, the relevant condition for the formation of chromite grains}. The solar values are shown as black dashed lines. The black boxes around the most anomalous grain 2\_37 represent a qualitative estimate of nuclear physics uncertainties as described in the text. The middle row is the same as the top row, except that the abundance of the radioactive \iso{53}Mn is not decayed into \iso{53}Cr.
    \label{fig:nittler_comp}
\end{figure*}

\begin{figure*}
    \includegraphics[width=\linewidth]{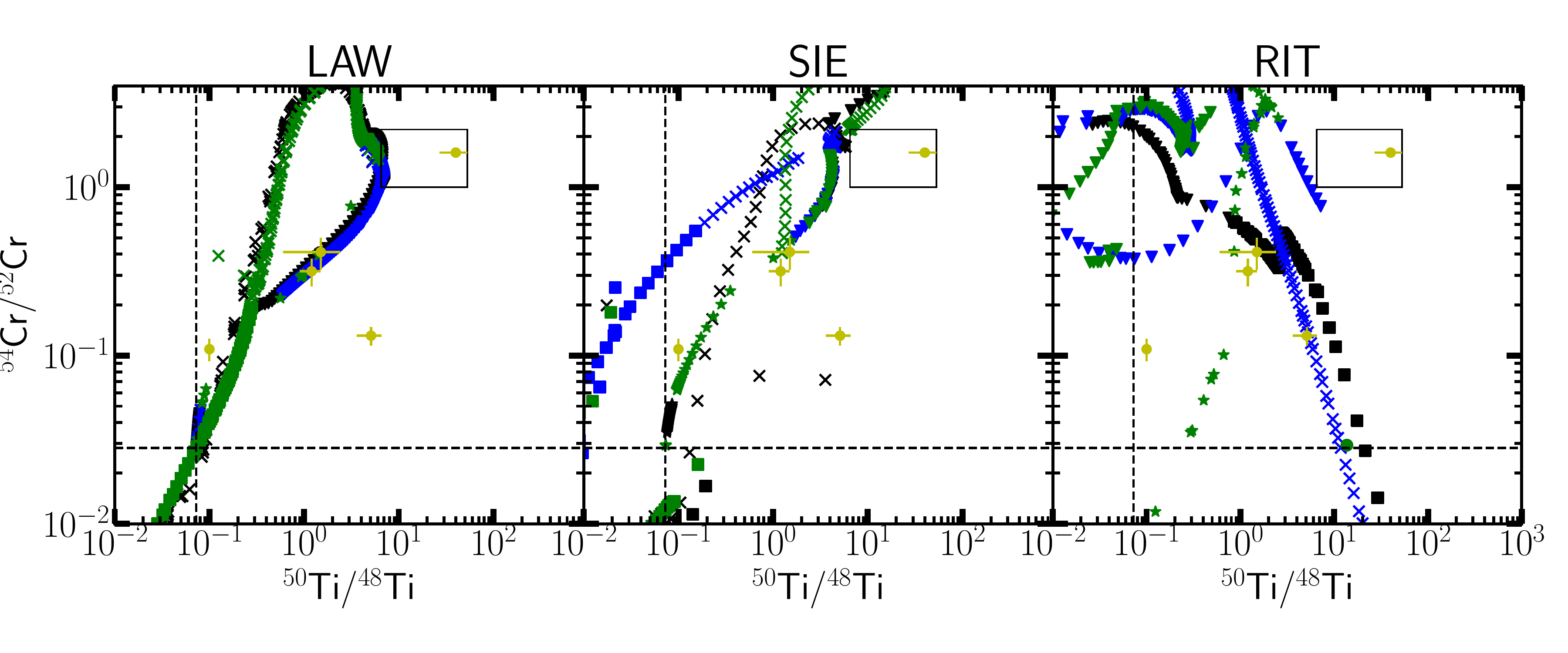}
    \caption{Comparison of the Cr isotopic ratio and Ti isotopic ratio as measured by \citet{2018ApJNittler} and as predicted isotopic ratios by the three CCSN data sets. The \iso{50}Ti/\iso{48}Ti ratio is shown instead of \iso{50}Cr/\iso{52}Cr, to investigate whether the mass 50 measurements are due to \iso{50}Ti or \iso{50}Cr. Colours and symbols are as in Figure~\ref{fig:nittler_comp}. The yellow \iso{50}Ti/\iso{48}Ti data points of \citet{2018ApJNittler} are calculated using the solar value for \iso{50}Cr/\iso{52}Cr. }
    \label{fig:nittler_comp50}
\end{figure*}

In Figures \ref{fig:nittler_comp} and \ref{fig:nittler_comp50} we compare the presolar chromite data of \citet{2018ApJNittler} to the three data sets of the CCSN model predictions. The predicted Cr ratios as shown in Figures~\ref{fig:nittler_comp} and \ref{fig:nittler_comp50} vary over orders of magnitude, as the different Cr isotopes are created and destroyed in different regions of the progenitor and its explosion (see figures in Appendix \ref{sec:app} and Table~\ref{tab:isotopes}). Each data point of the CCSN data sets in Figures \ref{fig:nittler_comp} and \ref{fig:nittler_comp50} corresponds to one numerical zone in a model, and we do not allow for mixing between zones. Only a few mass shells within each model can reach the stardust data points and are also O-rich (symbols in the figures), the condition necessary to form the chromite grains, as opposed to C-rich (not shown in the figures). We focus our discussion on finding a possible region that has a composition that matches the grain 2\_37 \citep{2018ApJNittler}, which has the most anomalous \iso{54}Cr/\iso{52}Cr and \iso{50}Cr/\iso{52}Cr ratios of the grains in this data sets. Less extreme values may be explained by dilution effects due to mixing with less processed material in the outer layers of the star, or with material in the interstellar matter \citep[ISM, see e.g., ][]{zinner:14}. First, we consider if a possible exact match of the models to the composition of 2\_37 exists, and second, we take into consideration in the discussion some of the uncertainties due to nuclear physics. These uncertainties are represented in Figures~\ref{fig:nittler_comp} and \ref{fig:nittler_comp50} by the boxes around grain 2\_37 and are described in detail below. 

We start by considering the top and bottom panels of Figures \ref{fig:nittler_comp}. For the three LAW models, the O-rich regions that can match the \iso{54}Cr/\iso{52}Cr ratio of 2\_37 are the He-ashes and C-ashes (triangles and crosses in the top left panel of Figure~\ref{fig:nittler_comp}). Between these two compositions, the He-ashes are located close to the required solar value of the \iso{53}Cr/\iso{52}Cr ratio, while the C-ashes provide a wide range of values for \iso{53}Cr/\iso{52}Cr. The C-ashes, however, match the \iso{50}Cr/\iso{52}Cr ratio of 2\_37, while the \iso{50}Cr/\iso{52}Cr ratio of the He-ashes are at least one order of magnitude lower than in 2\_37 (bottom left panel of Figure~\ref{fig:nittler_comp}). Note, that while the He-ashes of the 20 M$_{\odot}$ match both the \iso{54}Cr/\iso{52}Cr and \iso{53}Cr/\iso{52}Cr ratio of the grains, the \iso{50}Cr/\iso{52}Cr is almost three orders of magnitude too low.

We reach the same conclusions when considering the three SIE models (middle panels of Figure~\ref{fig:nittler_comp}), although the match is slightly worse than for the LAW models. When the \iso{54}Cr/\iso{52}Cr ratio is matched in the He- and C-ashes, the \iso{53}Cr/\iso{52}Cr ratio is at least 50\% higher than the solar ratio. When the \iso{53}Cr/\iso{52}Cr ratio in the C-ashes is equal to the solar ratio, the \iso{54}Cr/\iso{52}Cr ratio is lower than observed in 2\_37. For the \iso{50}Cr/\iso{52}Cr ratio, as in the case of LAW, the 2\_37 data point can only be reached in the C-ashes. The Ne-ashes of the 20 M$_{\odot}$ model reach values larger than in 2\_37 in the \iso{50}Cr/\iso{52}Cr (3 times larger than in 2\_37) and \iso{53}Cr/\iso{52}Cr ratio (4 times larger than in 2\_37). \iso{54}Cr/\iso{52}Cr on the other hand is at least 2 times lower than in 2\_37.

The Cr yields of the RIT models differ significantly from LAW and SIE (see Section \ref{sec:resultYields}). The C-, He- and H-ashes of the RIT 25 M$_{\odot}$ model reach the 2\_37 \iso{54}Cr/\iso{52}Cr ratio, but only for the He- and H-ashes is the \iso{53}Cr/\iso{52}Cr ratio also matched. None of the mass shells in this model reach the \iso{50}Cr/\iso{52}Cr ratio of 2\_37. The C-ashes of the 20 M$_{\odot}$ model reach the 2\_37 \iso{54}Cr/\iso{52}Cr ratio, while the \iso{53}Cr/\iso{52}Cr ratio is two times higher than the solar value observed in the grain. As for the \iso{50}Cr/\iso{52}Cr ratio the He-ashes within the 20 M$_{\odot}$ model have a composition very close to the grains, while the other parts of the ejecta do not. In the case of the 15 M$_{\odot}$ model we do not find any region that match the grain Cr composition.

As mentioned in Section~\ref{sec:dec_abu}, in the figures so far we have always presented results for abundances after radioactive decay. However, to exclude the radiogenic contributions can make a difference in the case of \iso{53}Cr because of the decay of \iso{53}Mn. The time between the CCSN and the formation of grains is currently unknown \citep[see e.g.,][]{2015Sarangi}. If the grains are created long enough after the explosion for all radioactive isotopes to decay and/or if the radioactive isotopes behave chemically in the same way as their daughter, then they are incorporated into the grains and contribute therein to the abundance of the stable isotopes and the decayed results apply. However, \iso{53}Mn has a relatively long half life of 3.74 Myr so it might be present in the grains, Mn is more volatile than Cr \citep{2003ApJLodders}, and Mn does not constitute as a major element in the spinel structure of the refractory Cr-oxide grains \citep{2010ApJDauphas}. Therefore, we also show in the middle row of Figure~\ref{fig:nittler_comp} the predicted isotopic ratios for the case that the dust grains have formed before \iso{53}Mn has decayed\footnote{None of the other isotopic ratios discussed here present an effect due to radiogenic decay, except for the \iso{57}Fe/\iso{56}Fe ratio, which only shows minor differences in the most central O-rich regions.}. When comparing the top and middle panels of Figure \ref{fig:nittler_comp} we see that the radiogenic contribution from \iso{53}Mn generally does not affect the ratios in regions that are relevant for the comparison to the grains, except for the Ne- and C-ashes in all the SIE models and the 20 \msun\ RIT model. In some of the mass shells within these regions, the non-decayed \iso{53}Cr/\iso{52}Cr ratios are smaller by factors of a few relative to the decayed values, and closer to 2\_37. 

\begin{figure*}
    \includegraphics[width=\linewidth]{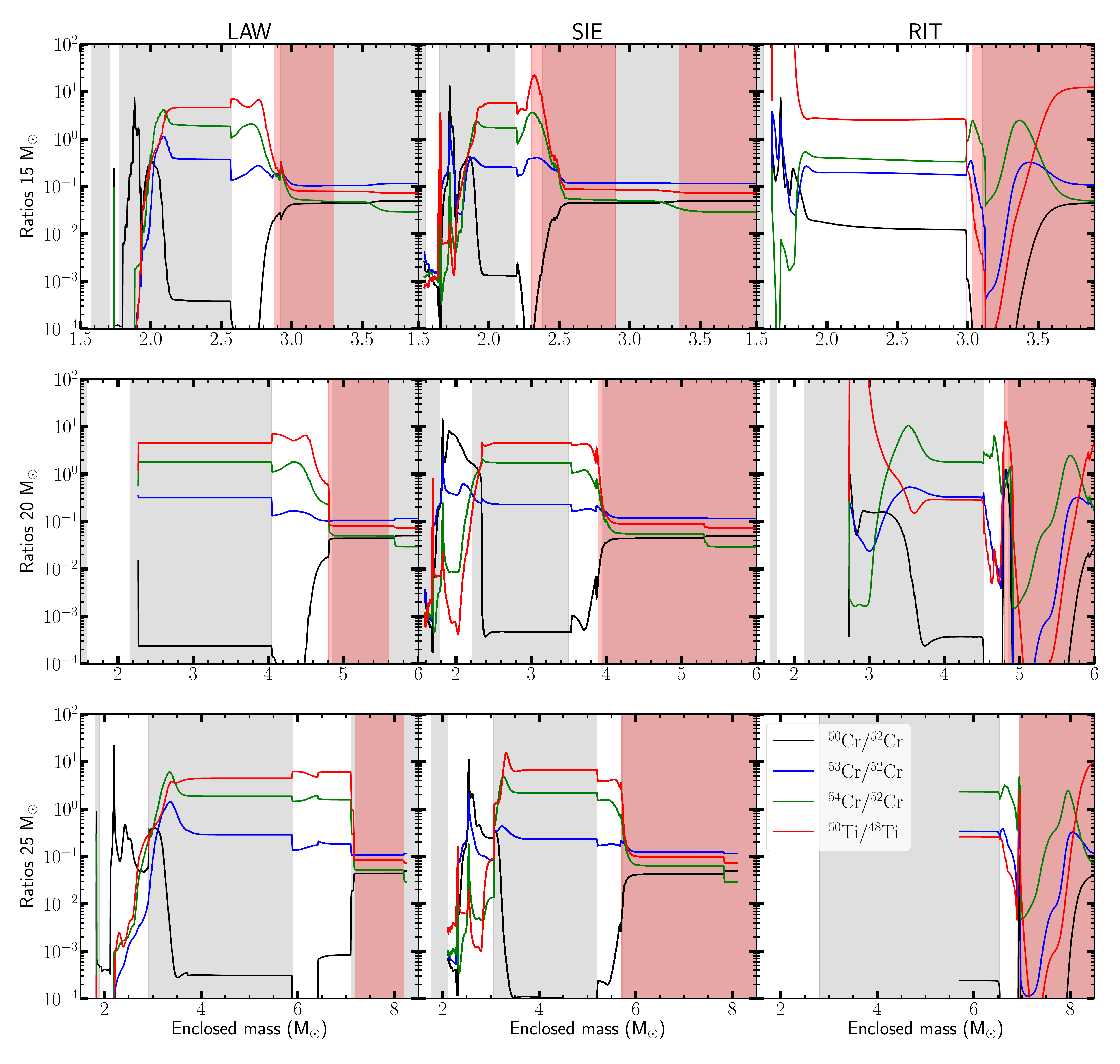}
    \caption{Cr and Ti isotopic ratios of all CCSN prediction (from top to bottom: 15 M$_{\odot}$, 20  M$_{\odot}$, and 25  M$_{\odot}$) after the explosion. The red regions are C-rich, and the grey and white bands indicate the different burning phases, following the nomenclature of Section \ref{sec:nomen}.}
    \label{fig:cr_ratios}
\end{figure*}

An ambiguity in the data of \citet[][]{2018ApJNittler} is that these authors were unable to distinguish between \iso{50}Cr and \iso{50}Ti at atomic mass A=50\footnote{The authors also checked for \iso{50}V and found that there is no V present in the grains}. Therefore, it is unclear whether the excess (above the solar ratio) of \iso{50}Cr/\iso{52}Cr ratio in the five grains, is due to \iso{50}Cr or \iso{50}Ti. For this reason,  we also compare our models in relation to the \iso{50}Ti/\iso{48}Ti ratio, see Figure~\ref{fig:nittler_comp50}. 

None of the three LAW models is able to reach the \iso{50}Ti/\iso{48}Ti ratio as found in 2\_37. However, the C-ashes of the 15 M$_{\odot}$ model and the He-ashes of the 20 M$_{\odot}$ model are only about a factor of three too low. The 15 M$_{\odot}$ SIE model reaches the 2\_37 value in its C-rich He-ashes. The other SIE models only approach the \iso{50}Ti/\iso{48}Ti of 2\_37, with either their He- or C-ashes. The shell merger region of the 15 M$_{\odot}$ RIT model is able to reach the \iso{50}Ti/\iso{48}Ti value of 2\_37, however, the \iso{54}Cr/\iso{52}Cr ratio in that region is at least two orders of magnitude less than in 2\_37. The 20 M$_{\odot}$ RIT model also approaches the \iso{50}Ti/\iso{48}Ti value of 2\_37, but again the \iso{54}Cr/\iso{52}Cr ratio is too low.

It is clear from the above analysis that it is impossible to identify large regions within the CCSN models that match three of the four isotopic ratios of the 2\_37 grain shown in Figures~\ref{fig:nittler_comp} and \ref{fig:nittler_comp50}. Therefore, we now try to identify regions that match the grain when taking into consideration uncertainties due to nuclear physics. In Figures~\ref{fig:nittler_comp} and \ref{fig:nittler_comp50}, the grain 2\_37 abundances are plotted within boxes. We defined these boxes as a reasoned qualitative estimate of the total uncertainty combining both the grain measurement error and the nuclear physics uncertainties affecting stellar model predictions, as discussed in more detail in Section~\ref{sec:cr_tests}. 

Specifically, the different axes of the boxes are set as follows: 
\begin{itemize}
    \item \iso{50}Cr/\iso{52}Cr: \iso{50}Cr is mainly created in explosive O- and Si-burning (see Table~\ref{tab:isotopes}), and these burning phases are typically not affected by nuclear physics uncertainties\footnote{See, for example, the sensitivity study by \citet{2013A&AParikh} for Type Ia supernovae, where these processes are active, which shows that \iso{50}Cr in not affected by reaction rates variations}. For the \iso{50}Cr/\iso{52}Cr box we therefore use the measurement of 2\_37 and its error bar: 0.317 $\pm$ 0.033. 
    \item \iso{50}Ti/\iso{48}Ti: the reaction rate tests in Section~\ref{sec:cr_tests} shows that this ratio can change from $\sim$4 to $\sim$15 when considering neutron-capture rate uncertainties (i.e., a factor $\sim$3.7) in the region close to the 2\_37 value (Figure~\ref{fig:ReactionTestsTi}). Therefore, we use this factor to extend the lower error bar of the lowest value of \iso{50}Ti/\iso{48}Ti in 2\_37, which is 27. The lower limit of the box is thus 7.2. For the upper limit of the box we use the upper error bar of the value of 2\_37. 
    \item \iso{53}Cr/\iso{52}Cr: we used the same reasoning as for \iso{50}Ti/\iso{48}Ti. The uncertainty factor resulting from the reaction rate tests in Section \ref{sec:cr_tests} is $\sim$2 (Figure~\ref{fig:ReactionTestsCr}). We use this factor to extend the upper error bar of 2\_37, as the model data is located at higher values. 
    \item \iso{54}Cr/\iso{52}Cr: for this ratio we extended both the upper and lower error bars of 2\_37, as the model data is located at both at higher and lower values than the value of 2\_37. The total variation is of a factor of $\sim$2 (Figure~\ref{fig:ReactionTestsCr}).
\end{itemize}
Note that for sake of simplicity we did not consider possible effects on \iso{52}Cr. This isotope is at denominator of all the isotopic ratios, therefore, changing its abundance would shift all the ratios by the same factor, resulting in a straight line passing through 2\_37, rather than a box. 

In Table~\ref{tab:oxides_summary}, we report the mass coordinates of the predicted model ratios which are located within the boxes as shown in Figures~\ref{fig:nittler_comp}-\ref{fig:nittler_comp50}, and using Figure~\ref{fig:cr_ratios} to identify the mass coordinates. We also list the ashes in which these mass coordinates are located, for both the decayed or non-decayed cases. Finally, we list the overlap regions considering mass 50 as either Cr or Ti, and these regions are indicated in Figure~\ref{fig:ratiosAlMg} with red dots. 

Table~\ref{tab:oxides_summary} shows that for all models we are able to identify a region in which the predicted ratio can be found within the box of \iso{54}Cr/\iso{52}Cr vs \iso{53}Cr/\iso{52}Cr. However, not all the models reach one of the other two boxes, which include an atomic mass 50 isotope. For five models an overlap between the \iso{54}Cr/\iso{52}Cr vs \iso{53}Cr/\iso{52}Cr mass range and at least one of the atomic mass 50 boxes can be found. The box around \iso{50}Ti is larger than the box around \iso{50}Cr due to the stronger sensitivity to reaction rate uncertainties. We find a higher number of overlap regions for the box around \iso{50}Ti (4) than for the box around \iso{50}Cr (2). We note that while an update of the reaction rates that affect the \iso{50}Ti/\iso{48}Ti ratio could lead to smaller boxes and thus a lower number of overlap regions, the location of these regions would not change. Specifically, the overlap regions that involve \iso{50}Cr/\iso{52}Cr are always located within the C-ashes, while for \iso{50}Ti/\iso{48}Ti they are located in the He-ashes and one in the C-ashes. Based on this analysis, the CCSN He-ashes and the C-ashes are both possible sites of origin for the Cr-rich grains. In the case of the C-ashes, it would most likely represent \iso{50}Cr. In the case of the He-ashes, the signal at atomic mass 50 would most likely represent \iso{50}Ti. This is in agreement with Table~\ref{tab:isotopes}, in which we show that \iso{50}Cr is produced in the C-ashes, while it is destroyed in the He-ashes. 

In all three LAW models we have identified overlap regions, as well as in the 15 and 25 \msun\ SIE models. We were unable to do this for the 20 \msun\ SIE model, likely because the temperature is higher in the region in the 20 \msun\ SIE model, where the \iso{54}Cr/\iso{52}Cr ratio falls within the box around 2\_37, than in the same region in the 20 \msun\ LAW model. In none of the RIT models were we able to identify overlap regions. In the 15 \msun\ model the reason for this is that the shell-merger and the explosive He-burning due to the temperature peak produce the Cr-isotopes in different ratios than in the other models. These two processes take place in the regions where the overlap is found in the LAW and SIE 15 \msun\ models. The 20 \msun\ RIT model experiences higher temperatures in the C- and He-ashes during the explosion, see Figure \ref{fig:temps}, also leading to different Cr-isotopic ratios. For the 25 \msun\ RIT model the main issue is the high mass cut, which excludes those regions in the ejecta where we find the overlap regions in the LAW and SIE models.

We also looked at the other models in the data set of Lawson et al (submitted), as shown as boxplots in Figure \ref{fig:isotope_boxplot}, which include a variety of values for the explosion energy and the mass cut. The 15 \msun\ models show little variability of the relevant isotopic ratios, while the 20 \msun\ models show differences in all isotopic ratios close to the mass cut. However, this region does not match the Cr isotopic composition of 2\_37. The variations in the 25 \msun\ models larger and present at more regions within the CCSN model. Most differences between the models, however, are small and fall within the uncertainty boxes in Figures \ref{fig:nittler_comp} and \ref{fig:nittler_comp50}, and therefore would not lead to more overlap regions than the ones already listed in Table~\ref{tab:oxides_summary}. The exception is that several 25 \msun\ models with high explosive energies provide a new overlap region as their \iso{50}Ti/\iso{48}Ti ratio reaches into the 2\_37 box within the He-ashes. This overlap region does not alter our findings that the isotope at atomic mass 50 is likely \iso{50}Ti in the He-ashes, and \iso{50}Cr in the C-ashes.

\begin{table*}
\caption{Mass coordinates (in M$_{\odot}$) at which the predicted ratio is within the boxes in Figures \ref{fig:nittler_comp} and \ref{fig:nittler_comp50}. The reported ranges cover solutions derived both for the decayed and non-decayed abundances. The overlap is defined as the overlap between \iso{54}Cr/\iso{52}Cr vs \iso{53}Cr/\iso{52}Cr and either \iso{50}Cr/\iso{52}Cr or \iso{50}Ti/\iso{48}Ti vs \iso{53}Cr/\iso{52}Cr. The location of the overlap is labelled by the ashes it is located in, and all overlap regions are indicated in Figure \ref{fig:ratiosAlMg}.}
\centering
    \begin{tabular}{cccc|ll}
         &  \iso{54}Cr/\iso{52}Cr vs & \iso{54}Cr/\iso{52}Cr vs & \iso{54}Cr/\iso{52}Cr vs & overlap with & ashes \\
                  &  \iso{50}Cr/\iso{52}Cr & \iso{53}Cr/\iso{52}Cr & \iso{50}Ti/\iso{48}Ti &  &  \\
         \hline
                  \multicolumn{5}{c}{LAW}\\
                           \hline
         15 M$_{\odot}$ & 2.02 & 2.02, 2.57 - 2.61, 2.74 - 2.78 & 2.58- 2.61, 2.77 - 2.78 & \iso{50}Cr: 2.02 & C \\
          15 M$_{\odot}$ & & & & \iso{50}Ti: 2.58 - 2.61, 2.77 - 2.78 &  He \\        
         20 M$_{\odot}$ & - & 4.05 - 4.50 & 4.05 - 4.19, 4.47 - 4.50 & \iso{50}Ti: 4.05 - 4.19, 4.47 - 4.50 & He \\
         25 M$_{\odot}$ & 3.12 - 3.15 & 5.88 - 7.10 & 6.43-6.44 & \iso{50}Ti: 6.43-6.44 & He \\
         \hline
         \multicolumn{5}{c}{SIE}\\
         \hline 
         15 M$_{\odot}$ & 1.87 - 1.88 & 1.87 - 1.89, 2.23 - 2.26 & 2.36 - 2.38 & \iso{50}Cr: 1.87 - 1.88  & C  \\
         20 M$_{\odot}$ & - & 2.28 - 2.34, 2.44 - 3.75 & - & - & - \\
         25 M$_{\odot}$ & - & 3.13, 4.80 - 5.57 & 4.80 - 5.20 & \iso{50}Ti: 4.80 - 5.20 & C (, He)\\ 
         \hline
         \multicolumn{5}{c}{RIT}\\
         \hline 
         15 M$_{\odot}$ & - &  3.04 - 3.06, 3.32 - 3.34 & - & - & - \\
         20 M$_{\odot}$ & 4.86 & 3.24 - 3.29, 5.61 - 5.64  & - & - & -\\
         25 M$_{\odot}$ & - & 6.54, 6.76 - 7.00  & - & - & - \\
    \end{tabular}
\label{tab:oxides_summary}
\end{table*}

The analysis above is based on comparison to the most anomalous grain 2\_37, and we justified this choice above by considering that less extreme values may be explained by invoking some dilution effect due to mixing with less processed material. However, it is interesting to check if the overall picture above would significantly change if we aimed at matching the two grains that are the second and third most anomalous in the \iso{54}Cr/\iso{52}Cr ratios. 

In the case of the LAW and SIE models, these grains could be matched by considering the C-ashes of the 15 and 25 \msun\ models or the He-ashes of the 20 \msun\ models, and the He-ashes 15 \msun\ for the LAW model. In this cases the atomic mass 50 is always only matched as \iso{50}Ti. The only difference between these two sets of solutions is that in the case of SIE, only the non-decayed abundances can match the two grains (as otherwise the addition of \iso{53}Mn produces too high abundance at atomic mass 53), while in the case of LAW both the decayed and the non-decayed predictions match the grains. Finally, these two grains can also be matched by the composition of the Ne ashes (shell merger) of the RIT 15 \msun\ model, in which case the isotope at atomic mass 50 could be either \iso{50}Cr or \iso{50}Ti. We note that all the solutions for all the three most anomalous grains reported in this section require relatively narrow CCSN mass regions. Other supernova studies that attempted to explain the presolar Cr-oxide data share the same problem of localised grain condensation \citep[see e.g.][]{2018ApJNittler,Jones2019b}. 

Furthermore, we note that as reported by \citet{2018ApJNittler} the \iso{57}Fe/\iso{56}Fe ratio of the grains is compatible within the error bar to the solar value except for one grain called 2\_81. The models can reproduce solar \iso{57}Fe/\iso{56}Fe ratios but only for very small specific ranges of mass coordinates, for example, in the O- and C-ashes at mass coordinates 2.2 and 2.9 \msun\, for the LAW 25 \msun\, model. This would make it very difficult for such signature to be predominant in the grains. However, the error bars on the \iso{57}Fe/\iso{56}Fe ratios are very large, of the order of the measured anomaly itself, because the overall abundance of Fe in the grains is very low (Larry Nittler, private communication). Therefore, we do not consider this as a strong constraint. 

\subsection{The $^{26}$Al signature in presolar C-rich and O-rich grains}
\label{sec:Al26signature}

\begin{figure*}
    \centering
    \includegraphics[width=\linewidth]{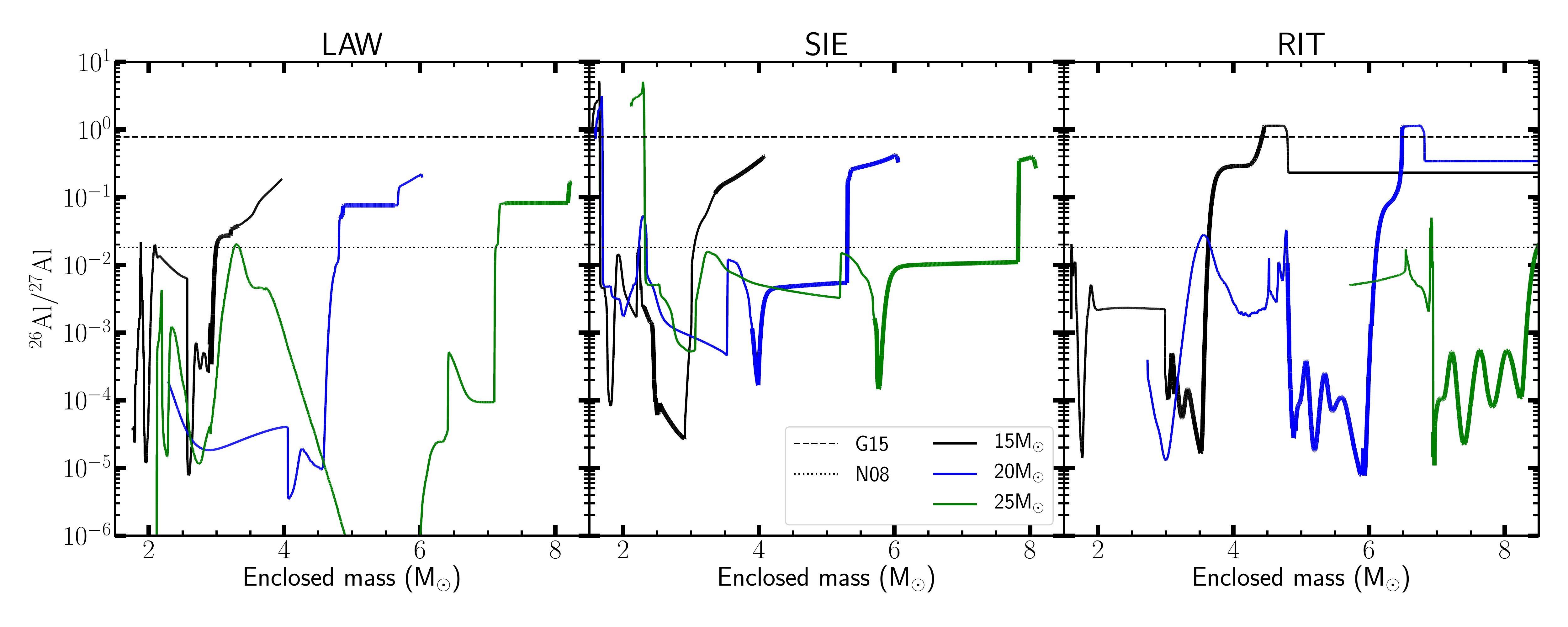}
    \caption{The $^{26}$Al/$^{27}$Al ratio is shown for the sets of stellar models, while the dashed line is the upper limit of SiC-X and Graphite data by \cite{2015ApJGroopman} and the dotted line the maximum limit of the four group 4 grains of \citet{2008ApJNittler}. The thick line segments indicate the carbon rich regions and the thin line segments the oxygen rich regions.}
    \label{fig:groopman}
\end{figure*}

In Figure~\ref{fig:groopman} we show the $^{26}$Al/$^{27}$Al ratio as predicted in the CCSN models. The dashed line indicates the highest values of the inferred initial $^{26}$Al/$^{27}$Al ratios, inferred from the Mg isotope composition of presolar SiC-X and graphite grains with CCSN origin \citep{zinner:14,2015ApJGroopman}. The dotted line represents the estimated initial $^{26}$Al/$^{27}$Al ratio of the Group 4 oxides that may also originate from CCSNe \citep{2008ApJNittler}.

None of the models of LAW in Figure~\ref{fig:groopman} reach the maximum ratio measured in \citep{2015ApJGroopman}, and only the Ne-ashes (see Figure~\ref{fig:cr_ratios} for identification of the ashes) of the SIE model with 25 M$_{\odot}$ initial mass reach an $^{26}$Al/$^{27}$Al ratio higher than the maximum measured in the stardust grains. The higher ratios are also reached even deeper in the ejecta of the 15 M$_{\odot}$ and 20 M$_{\odot}$ SIE models. However, these regions of the ejecta are not C-rich and have a very low absolute Al abundance. 
Therefore, including these layers in any realistic mixing of stellar material coming from different regions of the CCSN ejecta would not affect the final Al ratio in the resulting mixture. The RIT models reach the maximum measured ratio in the H-burning ashes that are mildly C-rich. Typical abundance signatures in C-rich grains from CCSNe, e.g. the enrichment in $^{15}$N and $^{28}$Si, and the $^{44}$Ca-excess due to the radiogenic contribution by $^{44}$Ti \citep[see e.g.,][]{amari:92, amari:95, besmehn:03} require some degree of mixing with other CCSN layers, where the $^{26}$Al enrichment is lower. It is still a matter of debate which components of the ejecta shape the mixtures observed in C-rich presolar grains. They could either undergo extensive mixing with deeper 
Si-rich regions \citep[e.g., from the so called Si/S zone][]{travaglio:99, 2018SciALiu}, or more localized mixing between C-rich layers  \citep[e.g.,][]{pignatari:13,pignatari:15, 2015ApJxu}. Some degree of contamination or mixing with isotopically normal material without $^{26}$Al has to be expected. More in general, the isotopic abundances from the RIT models would need to be compared directly with single presolar grains, to check if the $^{26}$Al enrichment can be reproduced along with other measured isotopic ratios \citep[e.g.,][]{liu:16,hoppe:18}. 

\cite{pignatari:15} showed that the ingestion of H in the He-shell of the massive star progenitor shortly before the onset of the CCSN explosion could potentially provide enough $^{26}$Al to reproduce the most $^{26}$Al-rich grains. None of the models considered in this work have developed late H ingestion events, and therefore we cannot fully explore the impact of these events in our study. While H-ingestion in CCSN models has been identified in stellar simulations since a long time \citep[e.g.][]{Woosley1995}, the quantitative impact of these events on the nucleosynthesis production is still poorly explored and there are large uncertainties. This is also due to the intrinsic difficulty of one-dimensional models to provide robust predictions for these events \citep[see, e.g., the discussion in][]{pignatari:15,hoppe:19}. We thus confirm that reproducing the high $^{26}$Al/$^{27}$Al ratios in C-rich grains is a still a major challenge for modern nuclear astrophysics. 

Nevertheless, based on previous works we can qualitatively expect that if H-ingestion and a following explosive H-burning take place, the neutron burst in the He-shell material will be mitigated compared to models without H-ingestion \citep[e.g.,][]{pignatari:15,2018liu_h_burning}. Therefore, the isotopes that are created in this region via neutron-captures relevant for this work, which include \iso{25,26}Mg and \iso{53,54}Cr (see Table~\ref{tab:oxides_summary}) as well as \iso{48,50}Ti, may be produced with smaller efficiency. In this case, the resulting nucleosynthesis might affect our overlap regions in Table~\ref{tab:oxides_summary}. We can speculate that the reduced \iso{53}Cr/\iso{52}Cr and \iso{54}Cr/\iso{52}Cr ratios could potentially affect the possibility for an overlap region to exist in the He-shell, depending on the exact remaining abundance of these isotopes. This aspect will need to be studied in the future, possibly using a new generation of massive star models informed by multi-dimensional hydrodynamics simulations of H ingestion \citep[e.g.,][]{clarkson:21}.

\citet{2008ApJNittler} concluded that 4 of their 96 analysed presolar oxide grains originated from CCSNe. The dotted line in Figure~\ref{fig:groopman} is the maximum of the $^{26}$Al/$^{27}$Al ratio of those four grains. All nine CCSN models shown in Figure~\ref{fig:groopman} reach this maximum value in an O-rich region. In the LAW models the dotted line is reached for the 15 M$_{\odot}$ in the C-ashes. The 20 M$_{\odot}$ and 25 M$_{\odot}$ model reach the dotted line in the He-ashes. The regions of the SIE models that reach the dotted line are for the 15 M$_{\odot}$ the He-ashes and the H-ashes, for the 20 M$_{\odot}$ model the inner C-ashes, and for the 25 M$_{\odot}$ the Ne-ashes. In the RIT models, the 15 M$_{\odot}$ model reaches the limit of \citet{2008ApJNittler} in the H-ashes, the 20 M$_{\odot}$ model in the C-ashes, and the 25 M$_{\odot}$ in the outer He-ashes. 

Therefore, in the case of these four presolar oxide grains that are assumed to have originated in CCSNe, there are extended O-rich regions consistent with the measured $^{26}$Al enrichment. Thus, local or more extended mixing of different stellar layers may potentially match the observed $^{26}$Al/$^{27}$Al ratio. The O isotopic ratios reported in \citet{2008ApJNittler} of these grains, however, are only reached in the envelope. Further analysis of all isotopic ratios obtained from these four grains is needed to conclude their region of origin. 

We note that while so far only the Group 4 oxides have been suggested to originate from CCSNe, recently \cite{hoppe:21} showed that some silicates from Group 1 and Group 2 could also be compatible with a CCSN origin, based on a comparison of their high $^{25}$Mg/$^{24}$Mg ratios to CCSN models affected by the H ingestion. More investigations are needed to define the full range of $^{26}$Al enrichment and $^{26}$Mg abundance signatures in all oxides and silicate grains with a possible CCSN origin.

\section{Discussion}
\label{sec:discussion}

\subsection{Effects of uncertainties in neutron-capture reaction rates on the Cr and Ti ratios}
\label{sec:cr_tests}

By considering three different data sets of stellar models, we have derived a qualitative estimate of the effect of stellar physics uncertainties and different computational approaches. This, however, does not provide us with a systematic way to check the effect of nuclear uncertainties. We have considered these separately and we present them here. 

As discussed in Section \ref{sec:creation_overv}, the main channel of production of $^{53}$Cr and $^{54}$Cr in regions where the chromite grains potentially originated from, are neutron captures on other Cr isotopes. The final abundances of \iso{53}Cr, \iso{54}Cr, \iso{48}Ti, and \iso{50}Ti after a given neutron flux episode are controlled mostly by their neutron-capture rates. To test how variations in these rates affect the Cr and Ti isotopic ratios, we preformed several dedicated tests using the MESA stellar evolution code, version 10398 \citep{2011ApJSPaxton,2013ApJSPaxton, 2015ApJSPaxton, 2018ApJSPaxton}. We used the settings for the massive star as described in \citet{Hannah2021} and considered models with an initial mass of 20\,M$_{\odot}$ with Z=0.014 evolved up to the core-collapse. We choose this progenitor model for our tests because the explosion has no significant impact on the abundances in the regions relevant for our analysis. The supernova explosion was not included in these tests, which is justified by the fact that the Cr isotopes in the C and He ashes are more significantly affected by the progenitor evolution than by the explosion, as shown in Appendix~\ref{sec:app}. 

We multiplied the neutron-capture reaction rates of interest by different constants, as indicated in Table~\ref{tab:tableCrmodeltest}. We choose variations in the direction that would help the models provide a better match to the most anomalous grain and we varied the rates by up to a factor of 2. This is larger than the up to 50\% uncertainty at 2$\sigma$ reported for the recommended values in the KaDoNiS database\footnote{See \url{https://kadonis.org/}} V0.2 \citep[][and therefore in the JINA reaclib database, which uses KaDoNiS]{2006AIPCDillmann}. However, these reactions were measured several decades ago: these current recommended values are from \citet{1977Kenny} for the Cr isotopes, from \citet{Allen1977} for $^{48}$Ti, and from \citet{Sedyshev1999} for $^{50}$Ti. Therefore it is possible that systematic uncertainties are much higher than the reported uncertainty.

\begin{table}
        \caption{Factors used to multiply the indicated reaction rates from their standard values in the 20 M$_{\odot}$ models considered in this section.}
            \centering
    \label{tab:tableCrmodeltest}
    \begin{tabular}{ccc}
      & $^{53}$Cr(n,$\gamma$)$^{54}$Cr & $^{54}$Cr(n,$\gamma$)$^{55}$Cr\\
      \hline          
      Model 1\footnote{Using the values of the KaDoNiS database \citep{2006AIPCDillmann}, which produces results very similar to those by LAW and SIE.} & 1   & 1   \\
      Model 2 & 1.5 & 1   \\
      Model 3 & 2   & 1   \\
      Model 4 & 1   & 0.5 \\
      Model 5 & 2   & 0.5 \\
      \hline 
              & $^{48}$Ti(n,$\gamma$)$^{49}$Ti &  $^{50}$Ti(n,$\gamma$)$^{51}$Ti\\
              \hline
      Model 6 & 2 & 1    \\
      Model 7 & 1   & 0.5 \\
      Model 8 & 2 & 0.5 \\
    \end{tabular}
\end{table}

\begin{figure}
    \centering
    \includegraphics[width=\linewidth]{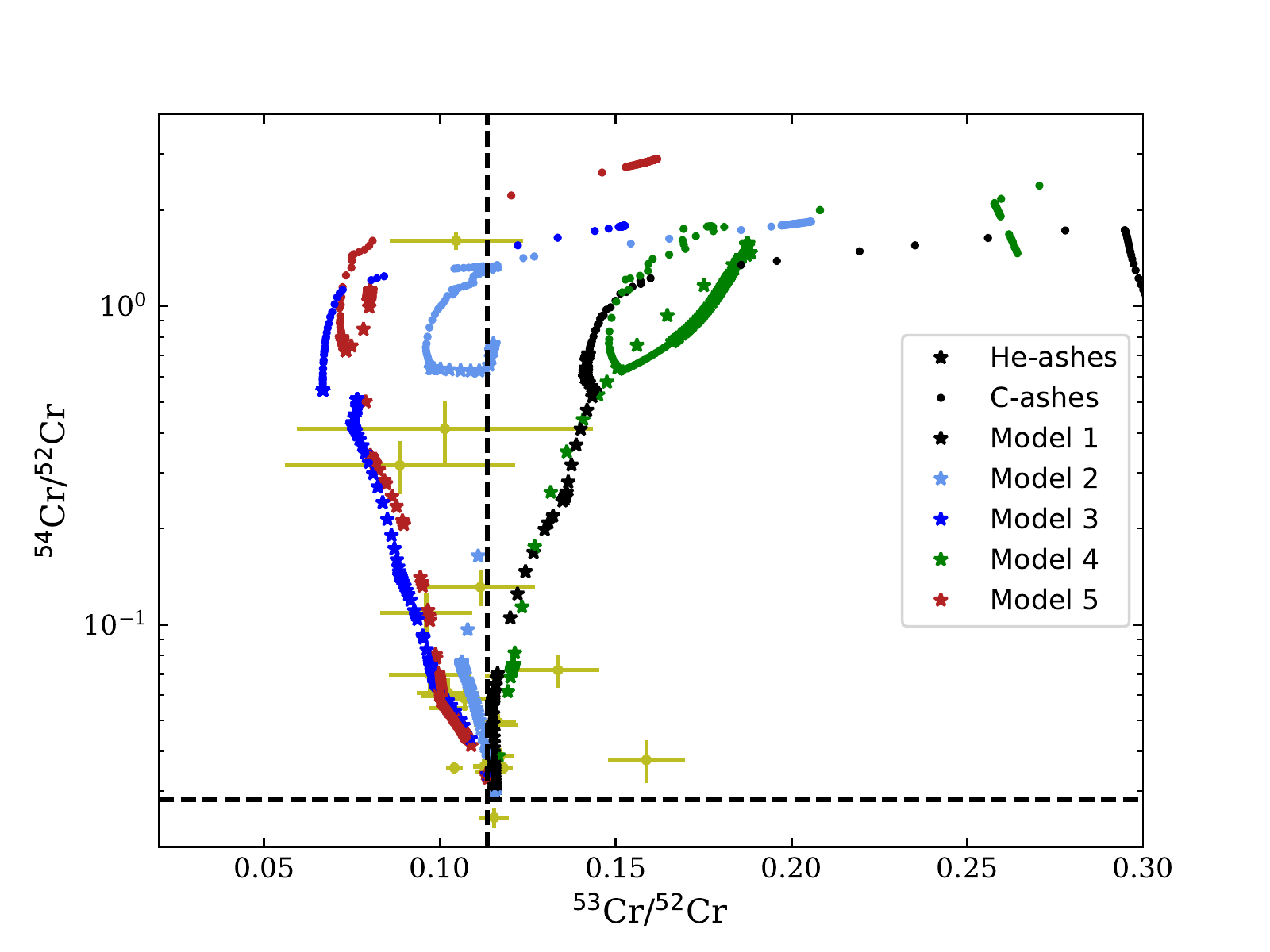}
    \caption{Cr-isotopic compositions resulting from the five 20 M$_{\odot}$ models calculated using different Cr neutron-capture rates, as listed in the top half of Table~\ref{tab:tableCrmodeltest}. The reference model is Model 1. As in Figure~\ref{fig:nittler_comp} the yellow points are the grains from \citet{2018ApJNittler}.}
    \label{fig:ReactionTestsCr}
\end{figure}

\begin{figure}
    \centering
    \includegraphics[width=\linewidth]{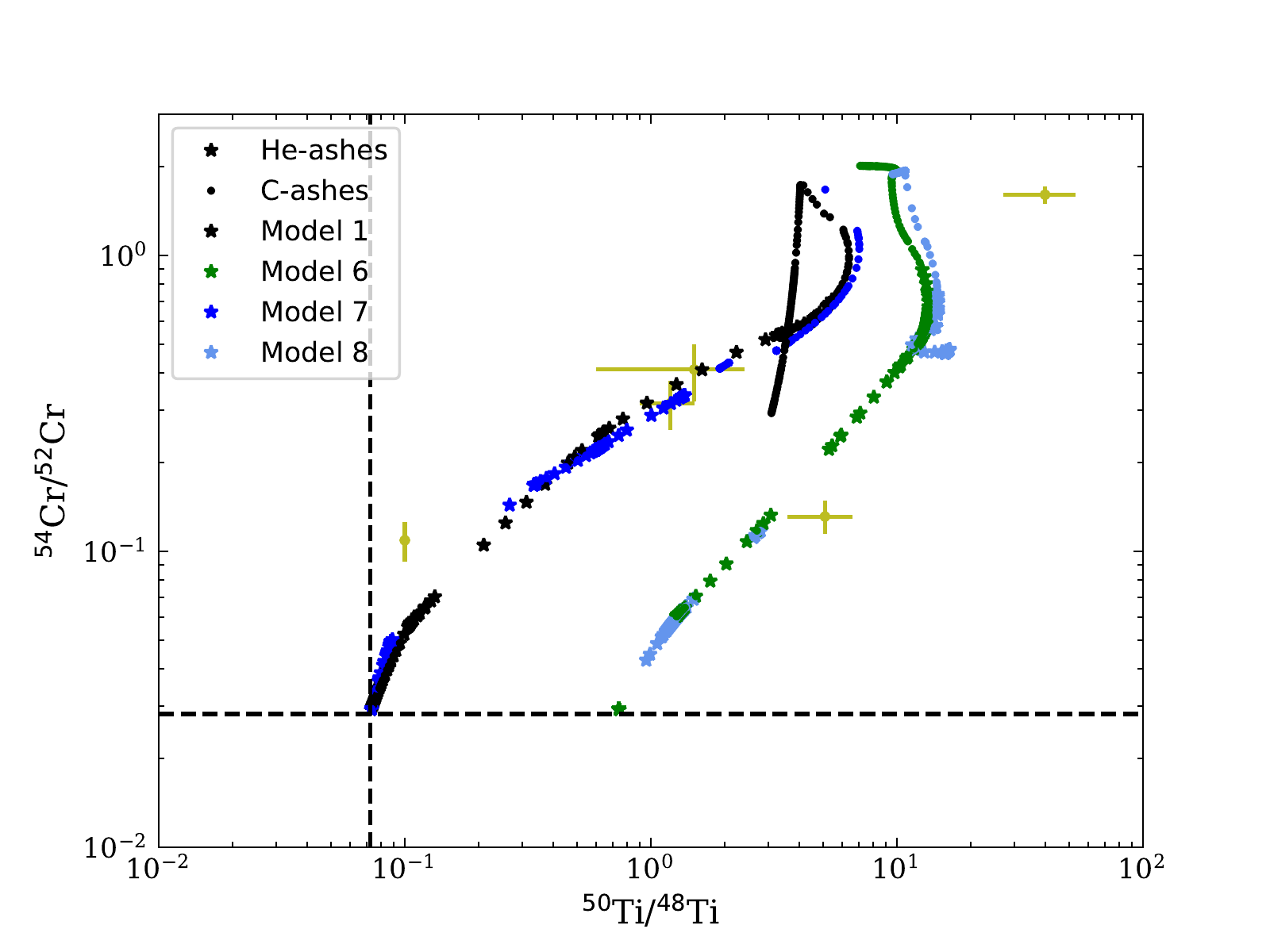}
    \caption{Same as Figure~\ref{fig:ReactionTestsCr} but for the $^{50}$Ti/$^{48}$Ti ratio resulting from the four 20 M$_{\odot}$ models calculated using different Ti neutron capture rates, and reference model Model 1. The rates used in these tests are listed in the bottom half of Table~\ref{tab:tableCrmodeltest}.} 
    \label{fig:ReactionTestsTi}
\end{figure}

Figures\,\ref{fig:ReactionTestsCr} and \ref{fig:ReactionTestsTi} show the results for the Cr and Ti isotopic ratios, respectively, in the C-ashes and He-ashes, which are the two possible sites of origin for the grains as described in Section~\ref{sec:resultsGrains}. In the case of the Cr isotopic ratios, two expected main trends are visible: (i) in the models with an enhanced $^{53}$Cr(n,$\gamma$)$^{54}$Cr rate only (Models 2 and 3) the $^{53}$Cr/$^{52}$Cr ratio decreases relative to the standard Model 1, for example from a maximum in the He ashes around 0.16 to a minimum of 0.07, i.e., roughly a factor of 2; (ii) in Model 4, with the reduced $^{54}$Cr(n,$\gamma$)$^{55}$Cr rate, the $^{54}$Cr/$^{52}$Cr ratio increases relative to Model 1, for example in the C-ashes from $\sim$1 to $\sim$2. In the combined test (Model 5), the $^{54}$Cr/$^{52}$Cr ratio increases further to $\sim$3 in the C-ashes compared to Model 1. Although these tests are only meant to provide a basic estimation of the impact of nuclear uncertainties, we can already derive that the uncertainties of the neutron-capture rates of Cr isotopes have a significant impact on stellar calculations. Therefore, new measurements of these neutron-capture rates are needed to reduce the uncertainty of the model predictions. 

When considering the results of the Ti tests, we find that increasing the $^{48}$Ti(n,$\gamma$)$^{49}$Ti reaction rate only (Model 6) leads to an increase of the $^{50}$Ti/$^{48}$Ti ratio. Decreasing the $^{50}$Ti(n,$\gamma$)$^{50}$Ti reaction rate only (Model 7) does not have a significant effect, because $^{50}$Ti is a magic nucleus and therefore has a very low neutron-capture cross section in both the two nuclear reaction setups. As a consequence, when both rates are changed in Model 8, the result is very similar to Model 6. We also tested the case for Ti with the rates multiplied and divided by 1.5 instead of 2, the results are very similar to those obtained by using the factor or 2. The result of these reaction rate tests concerning the Cr and Ti isotopes are used to define the boxes in Figures~\ref{fig:nittler_comp} and \ref{fig:nittler_comp50} as described in Section~\ref{sec:chromite}.

We did not test the impact of the nuclear uncertainties affecting the production of neutrons. In both the He-ashes and the C-ashes the $^{22}$Ne($\alpha$,n)$^{25}$Mg reaction is the main neutron source. The impact of its present uncertainty on He-burning and C-burning nucleosynthesis is well studied \citep[e.g.,][]{kaeppeler:94,heger:02,2010ApJPignatari}. A more precise definition of the competing $\alpha$-capture rates 
$^{22}$Ne($\alpha$,n)$^{25}$Mg and $^{22}$Ne($\alpha$,$\gamma$)$^{26}$Mg at relevant stellar temperatures is an open problem of nuclear astrophysics and an active line of research for many years  \citep[e.g.,][]{longland:12,talwar:16,2021PhRvCAdsley}.

\subsection{Al and Mg composition of the CCSN regions as candidate sites of origin of the chromite grains}
\label{sec:almg}

\begin{figure*}
    \centering
    \includegraphics[width=\linewidth]{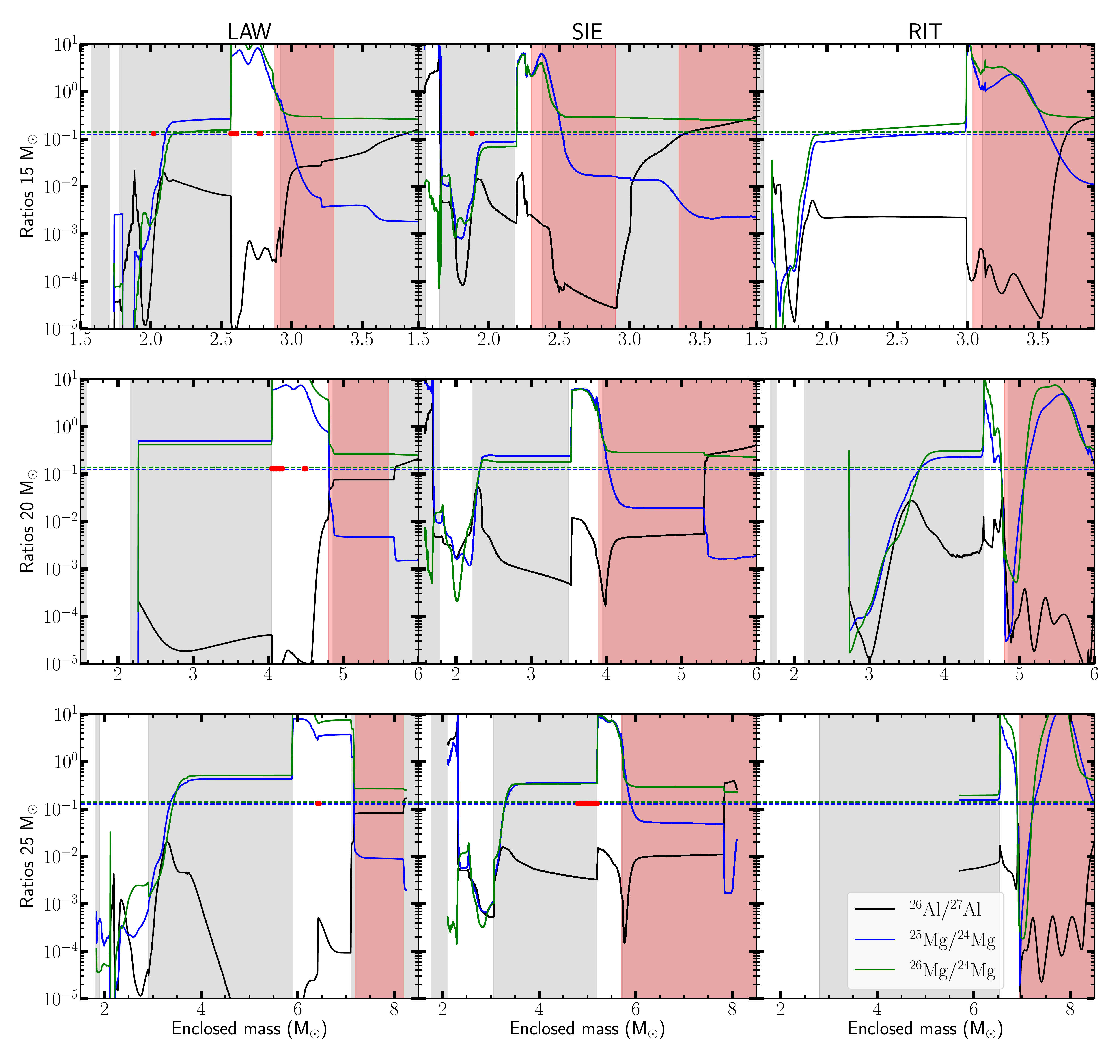}
    \caption{Ratios of \iso{26}Al/\iso{27}Al, \iso{25}Mg/\iso{24}Mg, and \iso{26}Mg/\iso{24}Mg (all decayed except for \iso{26}Al) for our three CCSN data sets after the explosion as function of the mass coordinate. The 15, 20, and 25 \msun\ models plotted in the top, middle, and bottom panels, respectively. The white, grey, and red bands represent the same regions as in Figure~\ref{fig:cr_ratios}. The blue and green dashed lines represent the solar values for the \iso{25}Mg/\iso{24}Mg and \iso{26}Mg/\iso{24}Mg ratios, respectively \citep[with values from][]{2011Bizzarro}. The red circles, plotted on the on the dashed lines for sake of visibility, correspond to the mass coordinates where the Cr and/or Ti isotopic ratios of the chromite grains by \citet{2018ApJNittler} match the ratios predicted by the models, as listed in the ``Overlap'' column in Table~\ref{tab:oxides_summary}.}
    \label{fig:ratiosAlMg}
\end{figure*}

Here we investigate the link between the Al and Mg isotopic ratios and the \iso{54}Cr enrichment in the chromite grains, because of the significance of the apparent correlation between \iso{54}Cr and \iso{26}Mg among planetary objects. As mentioned in the Introduction, our method in this and in the following subsection is valid only under the assumption that Al and Mg abundances are carried in the chromite grains and/or similar carriers enriched in Al and produced in the same region of the chromite grains. While the volatility of Cr and Mg in an O- and Cr-rich CCSN environment is poorly constrained, at least under early Solar System conditions they might be comparable: both elements start condensation in the spinel phase and their major host phases, although different, have similar 50\% condensation temperatures, \citet{2003ApJLodders}.

In Figure~\ref{fig:ratiosAlMg} we show the \iso{26}Al/\iso{27}Al, \iso{25}Mg/\iso{24}Mg, and \iso{26}Mg/\iso{24}Mg ratios as function of the mass coordinate of the three CCSN data sets. We also highlight the overlap regions listed in Table~\ref{tab:oxides_summary} as red dots, which represent the stellar zones where the Cr-composition of the chromite grains is matched. The Al and Mg isotopic ratios at the location of red dots in Figure~\ref{fig:ratiosAlMg} are therefore expected to reflect the nucleosynthetic signature of these two elements in the chromite grains. This nucleosynthetic signature may also allow us to determine if the excess at atomic mass 26 observed in the Solar System material to accompany the \iso{54}Cr excess \citep{2011ApJLarsen} is due to a \iso{26}Al and/or a \iso{26}Mg excess. 

As discussed in detail in Section~\ref{sec:chromite}, the two main regions of interest for the origin of the chromite grains are the C- and the He-ashes, which is where the red dots in Figure~\ref{fig:ratiosAlMg} are located. Specifically, for the LAW set, these are the He-ashes and the centre of the C-ashes in the 15 \msun\, model, the He-ashes in the 20 \msun\, model, and the inner C-ashes in the 25 \msun\, model. For the SIE models, the red dots are location in the centre of the C-ashes in the 15 \msun\, model, and the inner C-ashes in the 25 \msun\, model. In the RIT models there are no overlap regions for the most anomalous grain. However, if we consider the second and third most anomalous grains, the region between 2 and 3 \msun\ for the enclosed mass in the 15 \msun\ RIT model provides a possible match. The composition of this region is similar to the C-ashes of the SIE 25 \msun\ model, therefore in the following we do not discuss it separately. 

We remind the reader that these CCSN mass regions appear to be relatively narrow as we identified them in Section~\ref{sec:chromite} by trying to match specifically the most anomalous observed presolar Cr-oxide grain, without mixing with material of a different composition. 
While there is observational evidence that the composition of the ejecta can be asymmetric \citep[e.g.,][]{hoflich2004}, mixing within the supernova remnants is still poorly understood. Studies of high-density graphite grains and SiC grains of Type X suggested that small scale mixing between different inner and outer region of a supernova must occur to explain nucleosynthetic signatures typical of the inner layer (such as the initial presence of radioactive \iso{44}Ti and excess in \iso{28}Si), together with signatures from the outer layers, such as the He shell \citep{travaglio:99,2007ApJYoshida}. However, \citet{pignatari:13} matched the grains without invoking this mixing with the composition that is produced by the effect of increasing the energy of the explosion on the He-shell. In addition, \citet{2020ApJSchulte} argue that the CCSN ejecta (especially the material coming from the inner most regions of the massive star) is too energetic to condense prior to mixing with the cold interstellar medium. 
At the location of the red dots in the He-ashes in Figure~\ref{fig:ratiosAlMg}, the Mg isotopic ratios are roughly a couple of orders of magnitude higher than their solar values, because \iso{25}Mg and \iso{26}Mg are produced by the operation of the \iso{22}Ne+$\alpha$ reactions. This means that even if some \iso{26}Al is present here, it will not influence the total sum of \iso{26}Mg and \iso{26}Al. Furthermore, \iso{26}Al is mainly destroyed in the He-ashes by the neutron capture reactions \iso{26}Al(n,p)\iso{26}Mg and \iso{26}Al(n,$\alpha$)\iso{23}Na. 
In the C-ashes, the Mg isotopic ratios are typically below their solar values in the inner part, and above solar in the outer part, with the switch being model dependent. In the SIE 15 \msun\, model, they are below their solar values in the whole C-ashes. This is due to the fact that \iso{24}Mg is one of the primary products of C burning, therefore the Mg isotopic ratios \iso{25}Mg/\iso{24}Mg and \iso{26}Mg/\iso{24}Mg decrease towards their solar values. Subsequently, in the inner part of the C-ashes during the explosion is \iso{24}Mg not only strongly produced, but also \iso{25,26}Mg are destroyed via proton captures leading to the production of \iso{26}Al. Most of the red dots in the C-ashes are located in the region of the C-ashes where the Mg isotopic ratios are below their solar values. The exception is the 25 \msun\ model of SIE where the red dots are located at mass coordinate 4.8-5.2 \msun, which corresponds to Mg isotopic ratios a factor of a few higher than their solar values. We note that these red dots are the only ones in the C-ashes that match the \iso{50}Ti/\iso{48}Ti ratio. 

In summary, the CCSN models predict \iso{54}Cr enrichment (as signalled by the presence of the red dots) together with stable \iso{25}Mg and \iso{26}Mg excesses in the He-ashes, while in the C-ashes, both \iso{25}Mg/\iso{24}Mg and \iso{26}Mg/\iso{24}Mg can be either higher or lower than their solar values. In the next section we compare these findings to planetary materials. We also take into consideration the possible radiogenic contribution of \iso{26}Al to \iso{26}Mg.

\subsection{Expected isotopic variations in planetary materials}

\begin{figure*}
\centering
    \includegraphics[width=0.49\linewidth]{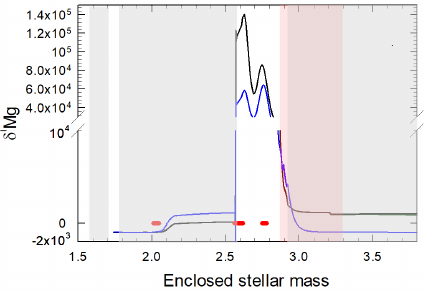}
    \includegraphics[width=0.49\linewidth]{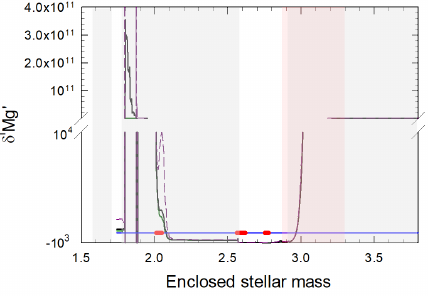} 
    \caption{For the LAW 15 \msun\ model we show: Left panel: the $\delta$\iso{25}Mg (blue), $\delta$\iso{26}Mg (black), and $\delta$\iso{26}Mg* (green) values (where \iso{26}Mg*=\iso{26}Mg+\iso{26}Al) assuming no mass fractionation of the CCSN ejecta. We also show the composition of an ejecta enriched in Al, $\delta$\iso{26}Mg*$_{\rm{E}}$ (purple dashed lines), where the Al/Mg ratio is set to 2, comparable to the abundances measured in colloidal presolar chromite grains \citep{2010ApJDauphas}}. The ratios are expressed as $\delta$-values, corresponding to their deviation to the terrestrial ratio in per mil (see text).
    Right panel: $\delta$\iso{26}Mg' and $\delta$\iso{26}Mg*', which are the internally normalised values by assuming maximum mass-dependent fractionation of the CCSN ejecta, i.e., the \iso{25}Mg/\iso{24}Mg ratio is set to the terrestrial value making $\delta$\iso{25}Mg' equal to 0. This natural mass-dependent fractionation is corrected for by the exponential mass fractionation law \citep[see, e.g.,][]{2011Bizzarro}.
    \label{fig:delta_Mg}
\end{figure*}

\begin{figure*}
     \includegraphics[width=0.5\linewidth]{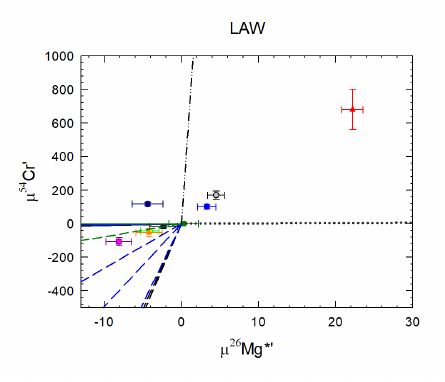}
     \includegraphics[width=0.5\linewidth]{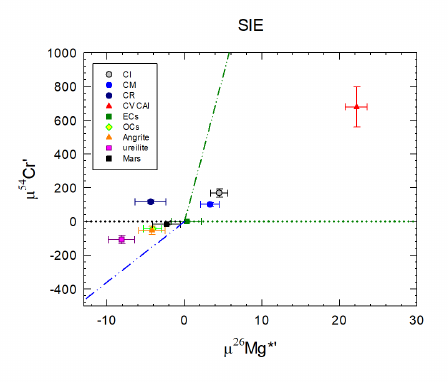}
     \includegraphics[width=0.5\linewidth]{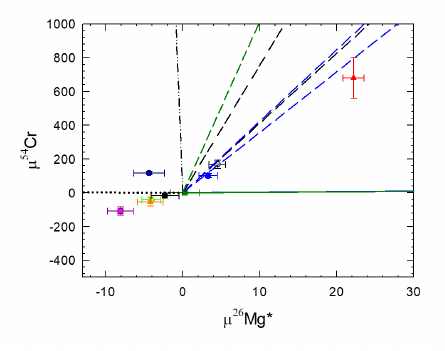}
     \includegraphics[width=0.5\linewidth]{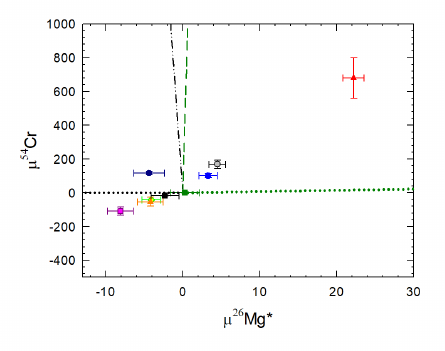}
     
\caption{$\mu$\iso{54}Cr' and $\mu$\iso{26}Mg*' values from meteoritic data \citep[colored symbols with 2$\sigma$ error bars, see legend, from][]{2011ApJLarsen} and predicted trajectories (lines) of $\mu$\iso{54}Cr' vs $\mu$\iso{26}Mg*' (top panels) and $\mu$\iso{54}Cr vs $\mu$\iso{26}Mg* (bottom panels) obtained by mixing between a solar component and the CCSN ejecta of the regions identified in Section~\ref{sec:chromite} (i.e., the red dots) for the LAW and SIE models (left and right panels, respectively)}. The colors of the mixing lines represent models for different stellar masses as before, i.e, black, blue, and green are used for the 15, 20, and 25 \msun, respectively. The C-ashes are indicated by the dotted lines, and the He-ashes by the solid lines. The dashed and dotted-dashed lines were calculated with Al- and Cr-enriched composition in the He and C ashes, respectively.
Top panels: The mass independent isotopic composition $\mu$\iso{54}Cr' and $\mu$\iso{26}Mg*' values of the CCSN ejecta were calculated by setting the \iso{52}Cr/\iso{50}Cr and \iso{25}Mg/\iso{24}Mg ratios to the NIST979 and DSM3 terrestrial standard values for Cr and Mg, respectively \citep{2011Bizzarro,2010GCAQin}. To follow the data reduction of meteorite measurements, we applied the exponential law to correct for mass fractionation \citep[][]{1978Russell}. 
Bottom panels: The CCSN ejecta are assumed to retain its isotopic composition, i.e., mass-dependent isotope fractionation is negligible. $\mu$\iso{54}Cr and $\mu$\iso{26}Mg* are calculated by simple deviation of their isotopic ratio values from the terrestrial standard values in ppm units without internal normalisation.
\label{fig:epsilon_Cr}
\end{figure*}

Here we compare the expected Cr and Mg isotopic compositions of the CCSN regions whose abundance composition match that of the chromite grains, as identified in Section \ref{sec:chromite}, to the Cr and Mg anomalies identified in planetary materials. We start with converting the predicted CCSN ejecta into commonly used variables in cosmochemistry. Then, we present mixing trajectories calculated with the CCSN ejecta and the solar composition, and compare these to the meteoritic data.

\subsubsection{Converting CCSN model data to normalized isotope ratios}
In the following we express the ratio of the abundance $N$ of isotope $i$ to isotope $j$ $(^{i}N/^{j}N)$ from models or measurements as part per mil ($\delta$) or per million ($\mu$) deviations from the terrestrial standard (TS):

\begin{equation}
\delta(\mu) ^{i}N = \left( \frac{(^{i}N/^{j}N)_{\rm sample/model}} {(^{i}N/^{j}N)_{\rm TS}} -1 \right) \times 10^3\, (\times 10^6).
\label{eq:delta_mu}
\end{equation}

Mg isotopic ratios of planetary materials are routinely measured with precise correction for instrumental mass-dependent fractionation (IMF) using the standard ``bracketing method'' \citep{2001Galy}. These IMF corrected values can be interpreted as the true values of the studied samples. They should reflect both the original nucleosynthetic mass-independent (which we indicate as $\delta$\iso{26}Mg$^{*}$) make-up of the analysed materials and all the physical processes that led to natural (as opposed instrumental) mass-dependent isotope fractionation of the sample during its chemical history. 

Unfortunately, the extent of the natural mass-dependent isotope fractionation, which we need to remove in order to obtain the original nucleosynthetic signature $\delta$\iso{26}Mg$^{*}$, is not precisely known \citep[see e.g.,][]{1977Wasserburg}. For meteorites and planetary samples, it is generally assumed that all the \iso{25}Mg/\iso{24}Mg deviation from the solar values as shown by the IMF corrected values is caused by natural mass-dependent fractionation. We note that the deviations from the solar \iso{25}Mg/\iso{24}Mg value are small (on \% level). The normalisation accounts for the maximum possible natural mass fractionation allowed by the data. The \iso{26}Mg/\iso{24}Mg ratio is therefore corrected for natural mass-dependent fractionation by using the exponential fractionation law \citep{2001Galy} and setting the \iso{25}Mg/\iso{24}Mg ratio to the terrestrial value. This is referred to as internal normalisation which results in a $\delta$\iso{26}Mg*',
identified as the remaining nucleosynthetic mass-independent anomaly. This is calculated as $\delta$\iso{26}Mg*'=$\delta$\iso{26}Mg - $\delta$\iso{25}Mg/$\beta$, where $\beta$ is the exponent of mass fractionation \citep[see e.g.][]{2011Bizzarro}. Finally, we note that the original mass-independent $\delta$\iso{26}Mg* (and $\delta$\iso{26}Mg*') should reflect both the contribution from the nucleosynthetic \iso{26}Mg and the production of radiogenic \iso{26}Mg by now extinct \iso{26}Al, which is also produced by nuclear reactions in the star. 

We note that in case of Cr, meteoritic and planetary data is obtained via thermal ionization mass spectrometry using internal normalisation, where instrumental and natural mass fractionation are corrected together and therefore cannot be distinguished. 

A problem arises when we wish to compare model predictions to meteoritic data and convert the CCSN yields to internally normalized $\delta$ values. The issue is that the stellar \iso{25}Mg/\iso{24}Mg or \iso{50}Cr/\iso{52}Cr ratios are almost never equal to the terrestrial values (see Figure~\ref{fig:ratiosAlMg}), and that some of the nucleosynthetic sites that can produce the chromite grains, the \iso{25}Mg/\iso{24}Mg ratios differ from their solar value by up to 3 orders of magnitude. Therefore, if we apply internal normalisation using the terrestrial \iso{25}Mg/\iso{24}Mg value to obtain the $\delta$\iso{26}Mg*' of the ejecta, we automatically imply that any deviation from the terrestrial value is due to mass fractionation, which is clearly not the case. There are two options to consider: (i) we apply internal normalization in order to treat the data the same way as in case of laboratory measurements \citep[see][]{2004Dauphas} or (ii) we take the model results as the true values of the ejecta and assume no natural mass fractionation, i.e., the isotope ratio used for normalization is not taken as the terrestrial value.

In Figure~\ref{fig:delta_Mg} we show the $\delta$-value representation of the Al and Mg isotopic ratios of the LAW 15 \msun\ model (see Figure~\ref{fig:ratiosAlMg}, left top panel) as an example. We show both $\delta$\iso{26}Mg (calculated only considering the contribution of \iso{26}Mg at atomic mass 26) and $\delta$\iso{26}Mg$^*$ (calculated considering the contributions of both \iso{26}Mg and \iso{26}Al at atomic mass 26), to highlight the impact of the abundance of \iso{26}Al on the total mass budget at atomic mass 26. We show the two options above to convert the CCSN ejecta: the $\delta$ values are calculated (i) in the right panel, assuming maximum mass fractionation by setting the \iso{25}Mg/\iso{24}Mg ratio to its terrestrial standard value \citep[DSM3 standard, see ][]{2011Bizzarro} and (ii) in the left panel, assuming no mass fractionation of the ejecta.  

This figure illustrates how the amplitude of isotope variations changes when using $\delta$ values instead of simple isotope ratios as in Figure~\ref{fig:ratiosAlMg}. Two main effects are visible: (i) when the \iso{i}Mg/\iso{24}Mg ratio is lower than the terrestrial value, the ratio in Eq. \ref{eq:delta_mu} becomes negligible and the $\delta$-value approaches $-10^3$ (the $\delta$-scale is not linear, see Eq. \ref{eq:delta_mu}), e.g., as in the mass range below 2.1 \msun\ in the left panel, and (ii) the internally normalised $\delta$\iso{i}Mg' values (i.e., when setting \iso{25}Mg/\iso{24}Mg to the terrestrial value) magnify anomalies with respect to the \iso{25}Mg abundance (right panel), as this is again a non-linear transformation of data because of the exponential fractionation law. Overall, the impact of \iso{26}Al at atomic mass 26 is not significant at the location of the nucleosynthetic sites of our interest (at the red dots, where the black and green lines overlap). 

For the 15 \msun\ LAW model, the only location in the star where there is a difference between $\delta$\iso{26}Mg' and $\delta$\iso{26}Mg*' is the inner C-ashes, where the strong depletion of \iso{26}Mg accompanied by the enhancement of \iso{26}Al generates a separation between the $\delta$-values calculated using \iso{26}Mg only (green line) or using \iso{26}Mg+\iso{26}Al (black line). However, no red dots are present in these regions of the 15 \msun\ LAW model, therefore, the \iso{54}Cr-rich grains are not matched here. A similarly strong contribution of the \iso{26}Al abundance relative to the \iso{26}Mg abundance at atomic mass 26 in the calculation of the $\delta$-values shown in Figure~\ref{fig:delta_Mg} only develops in the Ne-ashes, at a mass coordinate of about 1.8 M$_{\odot}$. We also checked the behaviour of the other models and found that the contribution of \iso{26}Al to atomic mass 26 also becomes relevant in the Ne-ashes for the 25 \msun\ LAW model and all SIE models, as well as in the shell-merger region of 15 \msun\ RIT model, i.e., in regions that did not produce the composition of the chromite grains. We found one candidate site in a more central region of the C-ashes in the 25 \msun\ SIE model which shows \iso{25,26}Mg/\iso{24}Mg ratios higher than the solar value, and while the \iso{26}Al production is ongoing, its abundance relative to \iso{26}Mg remains insignificant.

In addition, we show an example of a more likely scenario, where we calculate an Al-enriched $\delta$\iso{26}Mg*$_{\rm{E}}$ (purple dashed lines in Figure~\ref{fig:delta_Mg}) using an Al/Mg=2 ratio, similar to the value reported by \citet{2010ApJDauphas}. This calculation better represents an ejecta rich in refractory oxide phases. 
We find that this enrichment does not play a significant role, except in the case of the C-ashes when the data is internally normalised (right panel of Figure~\ref{fig:delta_Mg}, where the purple dashed line peaks at around 2.05 \msun). In the regions of interest here (the red dots), instead, the maximum contribution of \iso{26}Al to the total mass at atomic mass 26 even in this enriched case corresponds to an increase of at most 50\%.

\subsubsection{Mixing trajectories}

In Figure~\ref{fig:epsilon_Cr} we show the predicted trajectories of two-component mixing between the particular sites of CCSNe identified in Section \ref{sec:chromite} (i.e., the red dots, which denote the ejecta whose abundance composition matched Cr-isotopic signature of the presolar chromite grains) and the solar material with solar Cr and Mg abundances and terrestrial isotopic composition.
For comparison, we also show the small, correlated mass independent Mg and Cr anomalies reported in several meteorites as internally normalised $\mu$\iso{26}Mg*' vs $\mu$\iso{54}Cr' values from \citet{2011ApJLarsen}. Note that the data sets on CR chondrules and CAIs are omitted because these materials are more heterogeneous, showing up to 100 ppm variation in the stable Mg isotopes, and a 5\% variation in the initial \iso{26}Al abundance (see e.g. \citealt{LUU2019} and \citealt{2020Larsen} for more details).

In general, each mixing line is a hyperbola that connects two ``end-members'': the solar isotopic composition, in the origin by definition, and the isotopic composition of the specific CCSN region fitting the chromite grain composition. The curvature (K) of the line is determined by the relative abundance of the normalising isotopes (\iso{52}Cr and \iso{24}Mg) in the ejecta compared to the Solar System value: K=(\iso{52}Cr/\iso{24}Mg)$_{\rm{solar}}$/(\iso{52}Cr/\iso{24}Mg)$_{\rm{CCSN}}$, see \citet{1978Langmuir,2004Dauphas}. 
Therefore, the line features are determined by both the isotopic composition ($\mu$ values in the plots) and the elemental composition, relative to the solar value of the CCSN ejecta. 
Note that the full lines are hyperbolas, but the plots are zoomed into the region of the meteoritic data, therefore, they appear as linear. It is common practice to plot the mixing trajectories as symmetric lines going through the solar/terrestrial value representing not only the addition but also the subtraction or ``unmixing'' of a nucleosythetic component. For clarity, here instead we only plot the mixing trajectories that result from addition of CCSN material to the solar abundances, to indicate the composition vectors towards the CCSN composition and to highlight model differences. To calculate the CCSN end-member we show results with and without mass-dependent fractionation, as outlined in the previous section. In the top panels of Figure~\ref{fig:epsilon_Cr} we show the trajectories derived from internal normalised model data (as in the right panel of Figure~\ref{fig:delta_Mg}), and in the bottom panels of Figure~\ref{fig:epsilon_Cr} we show the trajectories derived from no mass fractionation in the CCSN ejecta (as in the left panel of Figure~\ref{fig:delta_Mg}).

The lines in the top and bottom panels are very different from each other because in both the C ashes and He ashes, the isotopic ratios that we use for internal normalization, \iso{50}Cr/\iso{52}Cr and \iso{25}Mg/\iso{24}Mg, are very different from the solar values. For example, in the He ashes, the \iso{50}Cr is completely destroyed and \iso{25}Mg is produced (see Appendix A). Therefore, these normalizing isotopic ratios are at least as anomalous than the isotopic ratios we are investigating (\iso{54}Cr/\iso{52}Cr and \iso{26}Mg$^{*}$/\iso{24}Mg). This leads to extreme transformation of the isotopic space when applying internal normalization, even resulting in a change of sign in $\delta$ and $\mu$ notation.

All the solid and dotted lines (corresponding to the C- and the He-ashes, respectively) are horizontal because both in the He- and C-ashes the abundance of Mg is much higher than the abundance of Cr, therefore, the signature of the Mg isotopic composition is stronger in this representation relative to that of the Cr isotopic composition (i.e., K is between 1 and 200). We also note that the Mg isotopic signature is always dominated by the stable Mg isotopes, rather than by \iso{26}Al. 

As mentioned in the Introduction and at the start of Section 5.2, there is no published study on Mg isotopes in presolar chromite grains that gives evidence that chromite grains carry anomalies in Mg isotopes. Nevertheless, \citet{2010ApJDauphas} reported elemental abundances in chromite grains showing variable enrichment of Al and Cr relative to Mg. Following this indication, we considered the possibility of CCSN material enriched in Al and Cr with respect to Mg, with respect to the CCSN calculated abundances. For simplicity here we made a test using Al:Mg:Cr=2:1:1 (dashed and dotted-dashed lines in Figure~\ref{fig:epsilon_Cr} for the He- and C-ashes, respectively), noting that this represents an enrichment for Al, because the Al/Mg ratios in the CCSN candidate site are $\simeq$0.15, higher than the solar ratio of 0.06, but still much lower than 2, and for Cr, because the Cr/Mg ratios in the CCSN candidate sites are even lower than the solar ratio of 0.01. In these enriched cases, the K value becomes lower than 0.1 and therefore the mixing lines deviate from horizontal.

In these enriched cases for the C-ashes, the mixing lines appear as almost vertical as they are dominated by the increased relative abundance of \iso{52}Cr in this part of the CCSN ejecta (see Appendix~\ref{sec:app}). None of them match the correlation displayed by the meteoritic data. This C-ashes material could still be in agreement with other interpretations of the measured planetary Mg isotopic data, suggesting an homogeneous proto-planetary disc for Mg isotopes on a level of a few ppm \citep[e.g.,][]{2008Jacobsen,Kita2013,LUU2019}. 

In the case of the enriched He-ashes, the \iso{52}Cr/\iso{24}Mg ratio relative to the solar ratio is less extreme and the highly variable $\mu$ values from the isotopic composition dominate the mixing trajectories. In the top panels in Figure~\ref{fig:epsilon_Cr}, where we use internal normalisation the trajectories match the meteoritic trend only with negative $\mu$ values. This solution, however, would be inconsistent with our assumption that the observed anomalies are carried by refractory CCSN grains because such grains would not be preferentially destroyed in the inner Solar System, relative to less refractory ISM dust. Therefore, their presence should result in positive anomalies.  
In the bottom panels instead, where we assume that the mass fractionation of the ejecta is negligible, the 15 and 20 \msun\ LAW models may generate the observed trend via mixing or unmixing of a positive CCSN component suggested by \citet{2011ApJLarsen}. Note that the composition of the SIE He-ashes is just outside the border of the box defined in Figure~\ref{fig:nittler_comp50}, therefore, they are not included in this plot, but if they did they would behave similarly to the LAW He-ashes.

\section{Conclusions}
\label{sec:conclusions}

We presented a detailed analysis of the production of the Al, Mg, and Cr isotopes from CCSN models with non-rotating, single star progenitors. We compared the total isotopic yields between seven CCSN data sets and we compared the isotopic composition as a function of mass coordinate from three CCSN sets to the isotopic composition measured in meteoritic stardust grains. We found potential nucleosynthetic origin sites of the chromite grains presented in \citet{2018ApJNittler}, and evaluated the contribution at atomic mass 26 at those potential sites. 

Concerning the total CCSN yields, we found that the seven CCSN data sets are mostly comparable to each other for the nine isotopes of interest: $^{24,25,26}$Mg, $^{26,27}$Al, and $^{50,52,53,54}$Cr. The main differences are due to different mass cuts (mainly driving variations in the abundances of Cr isotopes), the occurrence of a shell merger in the 15 \msun\ RIT models, and structural differences in the progenitors between the LAW and SIE data sets. Based on a detailed analysis of the production sites of these isotopes in the different models, we are confident that our findings are representative of most 1D CCSN models of solar metalicity. 

We compared the CCSN models to the composition of the chromite grain most anomalous in \iso{54}Cr: 2\_37, including an estimate of the uncertainties due to neutron capture rates on \iso{53}Cr, \iso{54}Cr, \iso{48}Ti, and \iso{50}Ti, based on our sensitivity tests.

For all models, we were able to identify mass regions within the CCSN ejecta where the \iso{54}Cr/\iso{52}Cr and \iso{53}Cr/\iso{52}Cr ratios are matched, however, the situation is more complicated for the ratios including the atomic mass 50 isotopes. Only in five out of nine models we could find a complete solution and only in small regions of the CCSN ejecta (see Table~\ref{tab:oxides_summary}). These solutions are all located in either the C-ashes or the He-ashes in all the models of LAW and in the 15 and 25 M$_{\odot}$ models of SIE. The three RIT models did not show any overlap regions with grain 2\_37. 
The regions that match the \iso{50}Cr/\iso{52}Cr ratio of 2\_37, and overlap with the regions matching its \iso{54}Cr/\iso{52}Cr and \iso{53}Cr/\iso{52}Cr ratios, are located in the C-ashes. In contrast, the regions that overlap with the \iso{50}Ti/\iso{48}Ti ratio are located in the He-ashes, with only one case in the C-ashes. 

When we consider the second and third most anomalous grains, we find again that both the C- and He-ashes from the LAW and SIE models can match these grains, however, in these cases the signal at atomic mass 50 must always come from \iso{50}Ti. Furthermore, when considering these two grains we also find that the shell-merger region in the 15 \msun\ RIT model could be a match, in which case the isotope at atomic mass 50 could be either \iso{50}Cr or \iso{50}Ti. 

We found that adding or not adding the radioactive \iso{53}Mn into \iso{53}Cr does not significantly affect the results. This is different from \citet{2019MNRASJones}, who found that for electron-capture supernova ejecta the partial inclusion of \iso{53}Mn in \iso{53}Cr is crucial to match the values of grain 2\_37.

We conclude that the chromite grains analysed by \citet{2018ApJNittler} could have originated from CCSNe. We emphasize that CCSNe are the most frequent stellar events among the production sites considered so far \citep[][]{2018ApJNittler,Jones2019b}, thus making them a likely candidates as the origin of these grains.

The Al-isotopic data from SiC-X grains are believed to have originated from CCSNe \citep{Groopman2015}. We confirm, however, that standard CCSN models do not produce enough \iso{26}Al in their C-rich ejecta to match these grains (see Figure~\ref{fig:groopman}). Therefore, mixing of layers within CCSN models and/or a proton ingestion into the He shell \citep{pignatari:15} are needed to match the grain data. 
For the few oxide grains of Group IV that also potentially originate from CCSNe \citep{2008ApJNittler}, we could match their Al-isotopic ratios, however, a multi-element isotope analysis is needed to evaluate the origin of those grains. 

In the candidate regions within the He ashes that reproduce the Cr isotopic ratios of the presolar chromite grains, the \iso{i}Mg/\iso{24}Mg ratios are orders of magnitude higher than the solar ratio. Here, \iso{26}Al is being destroyed by neutron capture reactions and has little effect on the total abundance at atomic mass 26. In the candidate regions within the C-ashes, the \iso{i}Mg/\iso{24}Mg ratios of the inner regions are orders of magnitude lower than the solar values, due to the production of \iso{24}Mg by carbon burning and the partial depletion of \iso{25}Mg and \iso{26}Mg by neutron capture and proton capture. While the abundance of \iso{26}Al is more significant at these latter site, it never dominates the production at atomic mass 26. Since presolar chromite grains alone may drive the variation of Cr isotopes in the proto-planetary disk, we conclude that the ejecta carrying the chromite grains could have also generated nucleosynthetic \iso{26}Mg isotopic variation in the disk, and such variation would be dominated by non-radiogenic, stable Mg isotopic anomalies.

We compared Cr and Mg isotopic anomalies measured in meteorites and planetary materials with our candidate sites from the CCSN models, and derived the expected mixing trajectories between the solar and the CCSN reservoirs under the simple assumption 
that such Al and Mg abundances are carried within the chromite grains and/or similar carriers enriched in Al. When considering CSSN refractory material enriched in Al and Cr relative to Mg, as suggested by the data of \citet{2010ApJDauphas}, we found that the ejecta of the C-ashes does not generate significant Mg isotopic variations due to the extreme \iso{52}Cr abundances relative to \iso{24}Mg.
The ejecta of the He-ashes, instead, can generate a trend similar to the apparent Cr versus Mg isotopic heterogeneity. This trend is positive, as required under our assumption of refractory carriers, only if the CCSN material is not double normalised. The validity of such a comparison method requires further investigation.

Future measurements of chromite grains with Resonant Ionisation Mass Spectrometry \citep[RIMS,][]{2016Stephan} are needed to identify if the signal at atomic mass 50 is related to \iso{50}Cr from the C-ashes or \iso{50}Ti from the He-ashes, since this instrument can in principle extract isotopes of a single element and avoid isobaric interference during the analysis. Furthermore, to study the link between stardust grains and planetary objects and investigate the possible \iso{26}Al heterogeneity of the proto-planetary disk, we have assumed here that Al and Mg abundances are also carried in the chromite grains and/or similar carriers enriched in Al. This assumption needs to be investigated via future Mg isotope study of the \iso{54}Cr-rich chromite grains. New measurements of the neutron-capture cross sections for the chromium and titanium isotopes are also required to improve the accuracy of CCSN predictions. Finally, supernova models based on self-consistent multidimensional simulations are needed to reduce the uncertainties that result from parameterized mass cuts and explosion energies. 

\section*{Acknowledgements} 
We thank Larry Nittler for discussion and for sharing unpublished Fe data. This research is supported by the ERC Consolidator Grant (Hungary) funding scheme (Project RADIOSTAR, G.A. n. 724560). We thank the ChETEC COST Action (CA16117), supported by the European Cooperation in Science and Technology, and the IReNA network supported by NSF AccelNet. This work was supported in part by the European Union ChETEC-INFRA (project no. 101008324). TL and MP acknowledge significant support to NuGrid from STFC (through the University of Hull's Consolidated Grant ST/R000840/1) and ongoing access to {\tt viper}, the University of Hull High Performance Computing Facility. A.S. acknowledges support from the U.S. Department of Energy through grant DE-FG02-87ER40328 (UM), Office of Science, Office of Nuclear Physics and Office of Advanced Scientific Computing Research, Scientific Discovery through Advanced Computing (SciDAC) program. Research at Oak Ridge National Laboratory is supported under contract DE-AC05-00OR22725 from the U.S. Department of Energy to UT-Battelle, LLC. MP thanks support from the National Science Foundation (NSF, USA) under grant No. PHY-1430152 (JINA Center for the Evolution of the Elements), and from the "Lendulet-2014" Program of the Hungarian Academy of Sciences (Hungary).

\bibliographystyle{aasjournal}
\bibliography{main,library}{} 

\appendix
\restartappendixnumbering

\section{Figures showing the production and destruction of the Mg, Al, and Cr isotopes of interest}
\label{sec:app}

\begin{figure}
\centering
    \includegraphics[width=\linewidth]{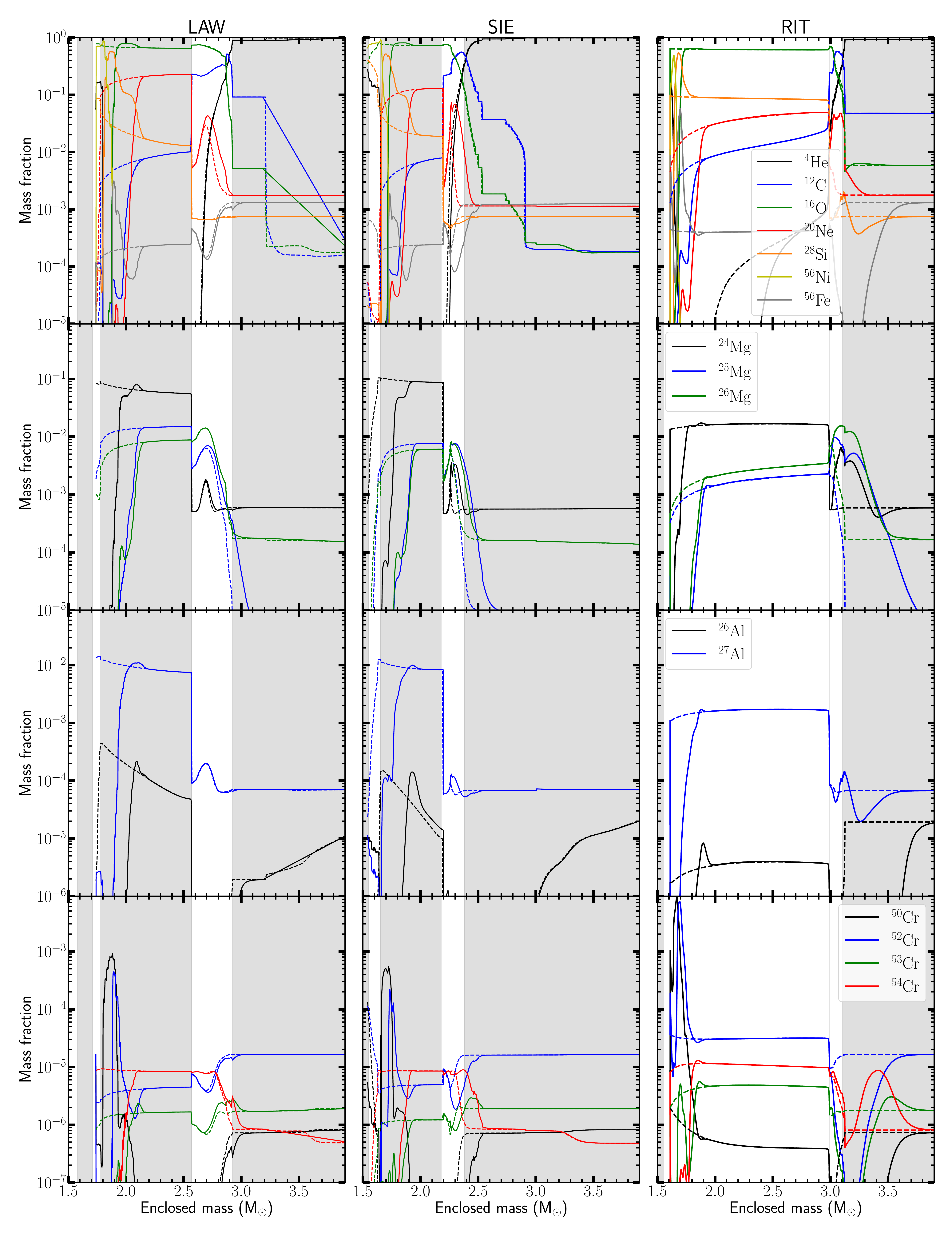} 
\caption{Non-decayed mass fraction profiles of the 15 M$_{\odot}$ models of the LAW, SIE, and RIT data sets. The dashed lines are the mass fractions in the progenitor, the solid lines show the fractions after the CCSN.} 
\label{fig:M15_spagh}
\end{figure}

\begin{figure}
    \includegraphics[width=\linewidth]{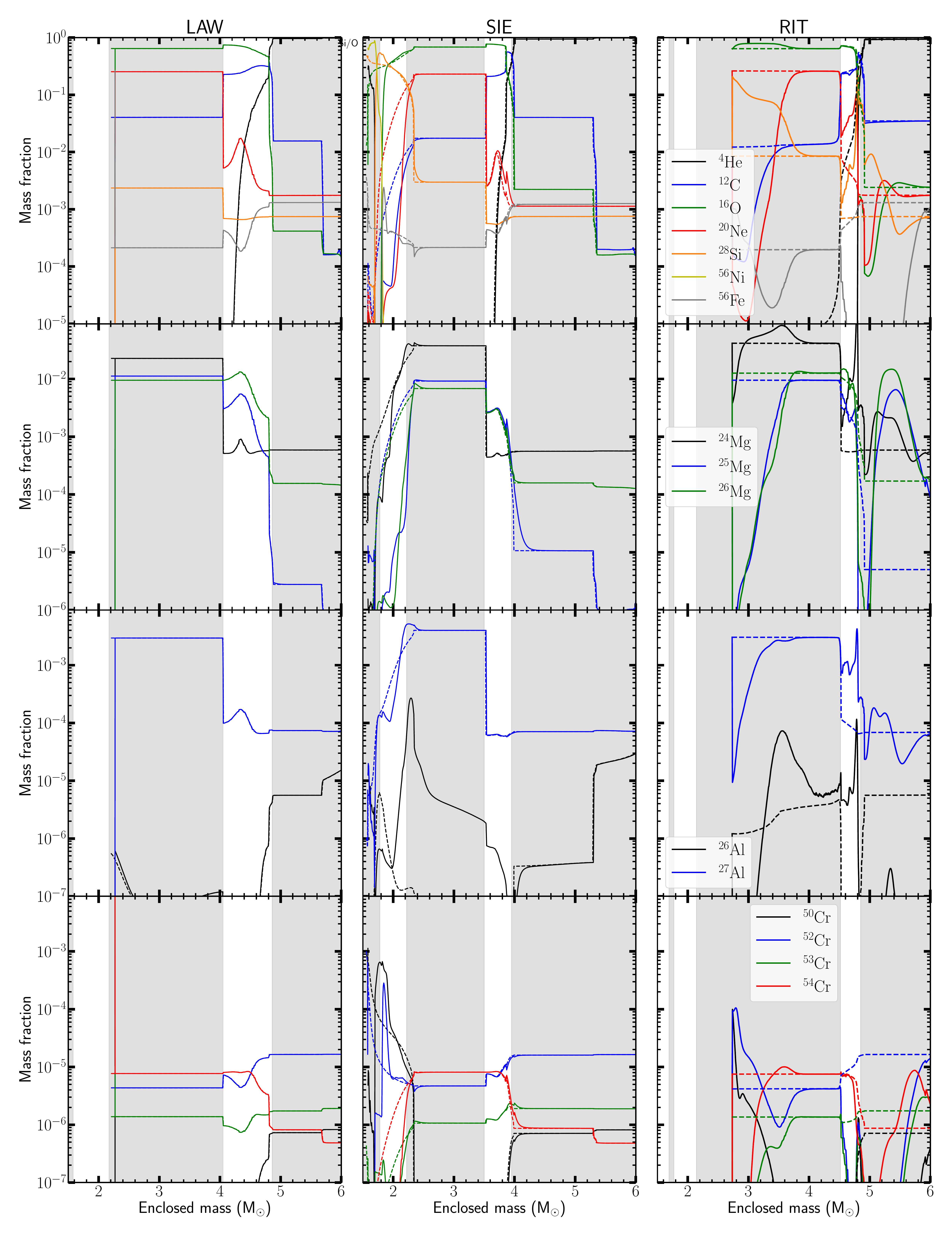} 
    \caption{Non-decayed mass fraction profiles of the 20 M$_{\odot}$ models of the LAW, SIE, and RIT data sets. The dashed lines are the mass fractions in the progenitor, the solid lines show the fractions after the CCSN.} 
\label{fig:M20_spagh}
\end{figure}

\begin{figure}
    \includegraphics[width=\linewidth]{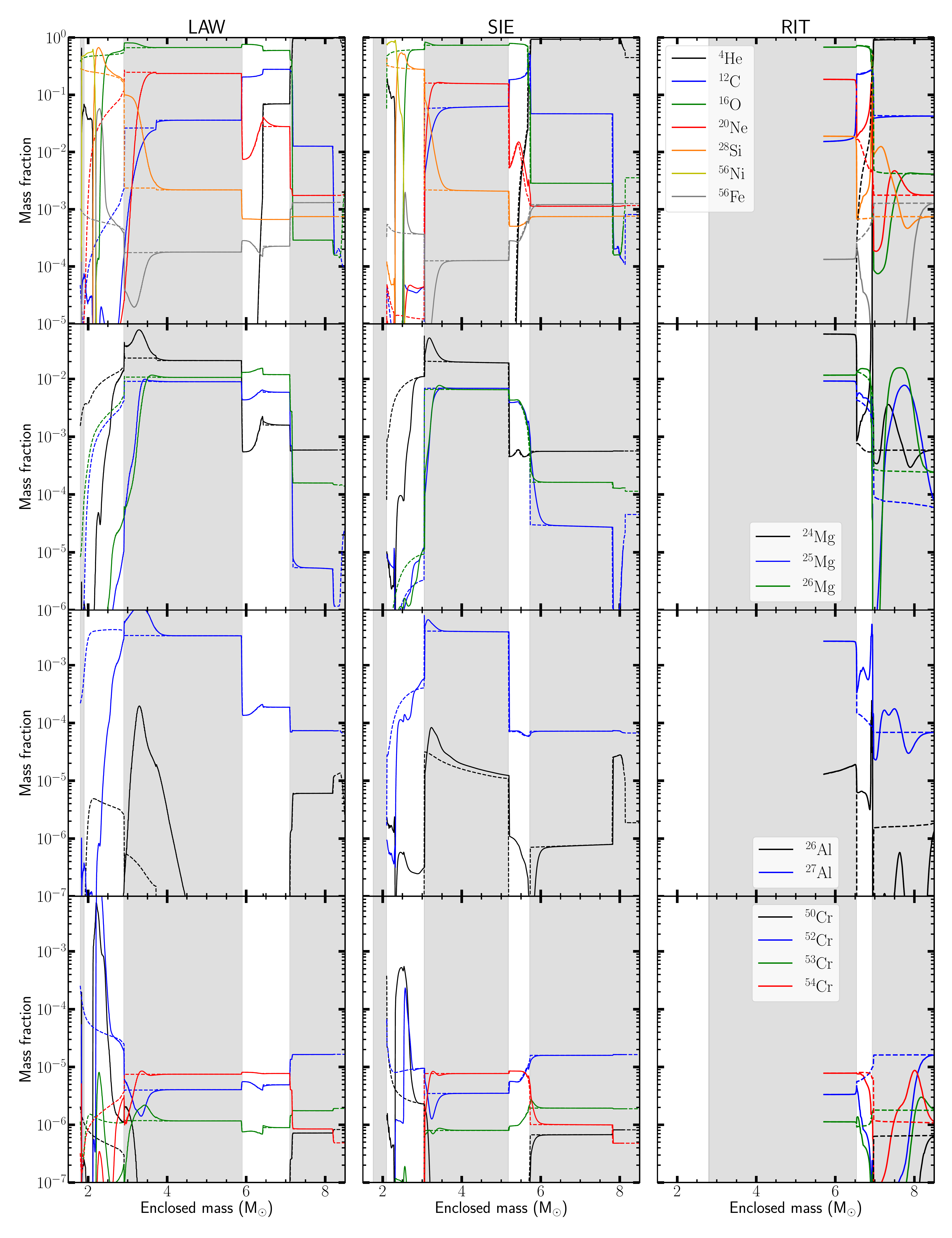}
    \caption{Non-decayed mass fraction profiles of the 25 M$_{\odot}$ models of the LAW, SIE, and RIT data sets. The dashed lines are the mass fractions in the progenitor, the solid lines show the fractions after the CCSN.} 
\label{fig:M25_spagh}
\end{figure}

\begin{figure}
    \begin{center}
    \includegraphics[width=\linewidth]{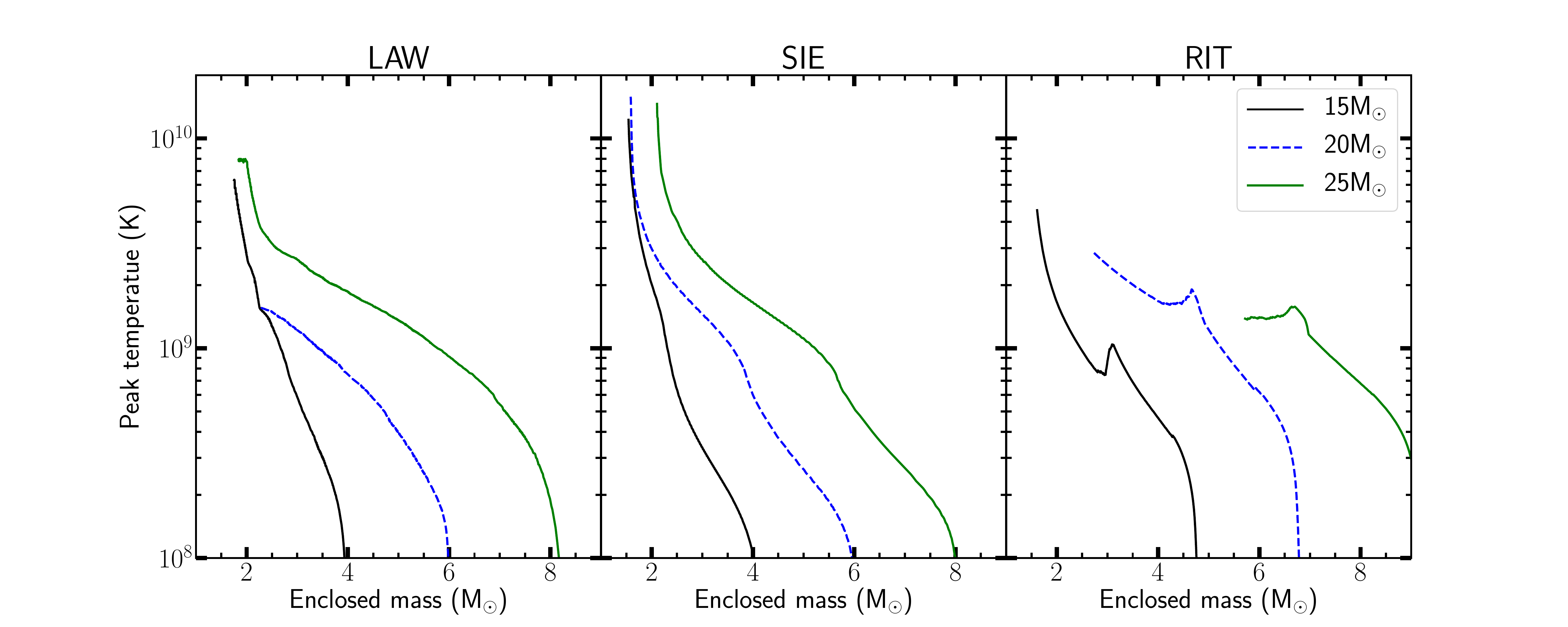}
    \caption{Temperature profiles for the explosions in all three CCSN model sets. The extra peaks in the RIT models (see e.g., in the 15 M$_{\odot}$ model at mass coordinate 3.2 M$_{\odot}$) are due to the analytic explosion model, which allows the velocity and thus temperature to increase when the shock decelerates \citep{2016ApJSPignatari}. These peaks allow for explosive He-burning in the H-ashes.}
    \label{fig:temps}
    \end{center}
\end{figure}

In this section we show in Figures~\ref{fig:M15_spagh}-\ref{fig:M25_spagh} the mass fractions of the three data sets of Lawson et al.(submitted), \citet{2018ApJSieverding}, and \citet{2018MNRASRitter} of the progenitor and the CCSN model as a function of mass coordinate. For each data set we show four figures for the three initial masses, being 15, 20 and 25 M$_{\odot}$: the top panel shows the stellar structure, and includes seven isotopes that allow us to identify the various burning phases within the progenitor and the explosion following the nomenclature as presented in Section \ref{sec:nomen}. The second panel shows the Mg isotopes, the third panel the Al isotopes, and the fourth panel shows the Cr isotopes as a function of mass coordinate. In Figure \ref{fig:temps} we show the temperatures in the CCSN models, which helps with identifying the differences in the nucleosynthesis between the nine CCSN models.

\end{document}